\documentclass[twocolumn,superscriptaddress,longbibliography,amsmath,amssymb,nofootinbib]{revtex4-1}
\usepackage{graphicx}
\usepackage[usenames,dvipsnames,svgnames,table]{xcolor}
\usepackage[colorlinks=true,linkcolor=RoyalBlue,citecolor=RoyalBlue]{hyperref}
\usepackage{soul}

\newcommand{\upperRomannumeral}[1]{\uppercase\expandafter{\romannumeral#1}}

\makeatletter

\usepackage{bm}

\newcommand{\bra}[1]{\langle #1 |}
\newcommand{\ket}[1]{|#1\rangle}
\newcommand{\bracket}[1]{\langle #1 \rangle}

\makeatother

\begin{document}

\title{Spin Nernst effect in the Paramagnetic Regime of An Antiferromagnetic Insulator}

\author{Yinhan Zhang}
\affiliation{Department of Physics, Carnegie Mellon University, Pittsburgh, Pennsylvania 15213, USA}

\author{Satoshi Okamoto}
\affiliation{Materials Science and Technology Division, Oak Ridge National Laboratory, Oak Ridge, Tennessee 37831, USA}

\author{Di Xiao}
\affiliation{Department of Physics, Carnegie Mellon University, Pittsburgh, Pennsylvania 15213, USA}

\date{\today}
\begin{abstract}

We theoretically investigate a pure spin Hall current driven by a longitudinal temperature gradient, i.e., the spin Nernst effect (SNE), in a paramagnetic state of a collinear antiferromagnetic insulator with the Dzyaloshinskii-Moriya interaction. The SNE in a magnetic ordered state in such an insulator was proposed by Cheng {\it et al.} [R. Cheng, S. Okamoto, and D. Xiao, Phys. Rev. Lett. 117, 217202 (2016)]. Here we show that the Dzyaloshinskii-Moriya interaction can generate a pure spin Hall current even without magnetic ordering. By using a Schwinger boson mean-field theory, we calculate the temperature dependence of SNE in a disordered phase. We also discuss the implication of our results to experimental realizations.
\end{abstract}
\maketitle

\section{Introduction}

Recent years have seen a surge of interest in issues related to spin transport in magnetic insulators. For practical purposes, the ability to transfer spin information in the absence of charge flow holds great potential for energy-efficient applications~\cite{uchida2008,uchida2010b,uchida2010,maekawa2011,bauer2012,ohnuma2013,kikkawa2013,chumak2015,cornelissen2015}. On the fundamental side, spin transport measurements can also provide valuable information about the ground state and low-energy excitations of correlated electronic systems~\cite{chen2013}.  In particular, a thermal Hall effect (THE) of spin excitations has been predicted~\cite{kastsura2010}.  In this effect, a longitudinal temperature gradient can drive a transverse heat current carried by charge-neutral excitations such as magnons or spinons. Since its prediction, the THE has been observed in a number of magnetic insulators~\cite{onose2010,hirschberger2015a,hirschberger2015b,ideue2017}, accompanied by extensive theoretical efforts~\cite{matsumoto2011a,matsumoto2011b,zhang2013,shindou2013,Lee2015,han2016,owerre2016,owerre2016b,owerre2017}. 
It is now recognized that, microscopically, the THE originates from nontrivial magnon dispersions due to
either chiral spin textures or non-symmetric spin-spin interactions, such as the Dzyaloshinskii-Moriya interaction (DMI).

However, in certain classes of magnetic insulators, the THE is symmetry-prohibited. Examples include magnetically disordered states at high temperatures and collinear antiferromagnets with combined time-reversal ($\mathcal T$) and inversion ($\mathcal I$) symmetry. For these systems, a spin Nernst effect (SNE) is symmetry-allowed nonetheless. In the SNE, spin currents with opposite polarization flow in the opposite transverse direction in response to a longitudinal temperature gradient.  As a result, the heat current vanishes, and we are left with a pure transverse spin current. The relation between the THE and the SNE is akin to the relation between the anomalous Hall effect and the spin Hall effect. The SNE has been predicted for magnets on a honeycomb lattice, either in antiferromagnets (AFM) below the N\'eel temperature in which the SNE is realized by magnons~\cite{cheng2016,zyuzin2016,lee2018}, or ferromagnets (FM) above the Curie temperature in which the SNE is realized by spinons~\cite{kim2016}.  Possible experimental signature of the SNE has also been reported in the antiferromagnetic insulator MnP$\rm{S}_{3}$ in the ordered phase~\cite{saitoh2017b}.

\begin{table}[b]
\center
\begin{tabular}{|c|c|c|}
\hline 
Collinear order & Ordered & Disordered\tabularnewline
\hline 
\hline 
FM & THE~\footnote{Refs.~\cite{owerre2016,kim2016}.} & SNE~\footnote{Ref.~\cite{kim2016}.} \\
\hline 
AFM  & SNE~\footnote{Refs.~\cite{cheng2016,zyuzin2016}.} & SNE~\footnote{This work.}
\tabularnewline
\hline 
\end{tabular}
\caption{Summary of the thermal Hall effect (THE) and the spin Nernst effect (SNE) in honeycomb magnets with a second nearest-neighbor Dzyaloshinskii-Moriya interaction.  Depending on the symmetry, the system exhibits either a THE or a SNE.}
\label{tab:table}
\end{table}

Actually, the honeycomb magnets can display either the THE or the SNE depending on their magnetic configurations, as summarized in Table~\ref{tab:table}.  The key ingredient here is a second nearest-neighbor DMI, which plays a similar role in spin transport as the spin-orbit interaction in electron transport.  In the ordered phase of a honeycomb FM, the broken time-reversal symmetry together with the DMI leads to the THE~\cite{owerre2016,kim2016}.  On the other hand, in both the disordered phase of the FM and the ordered phase of the AFM, the vanishing magnetization forbids the THE, but the DMI still allows the SNE~\cite{kim2016,cheng2016,zyuzin2016}. These results strongly hint that the SNE should also exist in the high-temperature disordered phase of the honeycomb AFM.

In this paper we present a detailed study of this effect using the Schwinger boson mean-field approach. We show that the SNE is indeed enabled by the DMI in the high-temperature disordered phase of a honeycomb AFM, and the transverse spin current is carried by the two pairs of conjugated spinon states connected by the combined $\mathcal{T}\mathcal{I}$ symmetry. Supplemented by a symmetry analysis, we calculate the reduced mean-field order parameters of the spinons, establish the disordered phase regime, and then identify the effect of a $\mathcal{T}\mathcal{I}$ conjugate pair on the pure SNE. Finally, we calculate the temperature dependence of the SNE coefficient in this disordered phase, and discuss its realization in real materials.

This paper is organized as follows.  In Sec.~\ref{Model_Method}, we introduce the honeycomb AFM model with a second nearest neighbor DMI, and present the mean-field solution to the Schwinger boson Hamiltonian.  This is followed by a discussion of the SNE in Sec.~\ref{Top_Spinons}, including its dependence on the temperature, the staggered field, and the DMI strength.  Finally, we comment on the limitations of our theoretical treatment and discuss possible material realizations of the SNE in Sec.~\ref{Disc}.

\section{Model and Method}
\label{Model_Method}

\begin{figure}[tb]
\centering
\includegraphics[width=1.0\columnwidth]{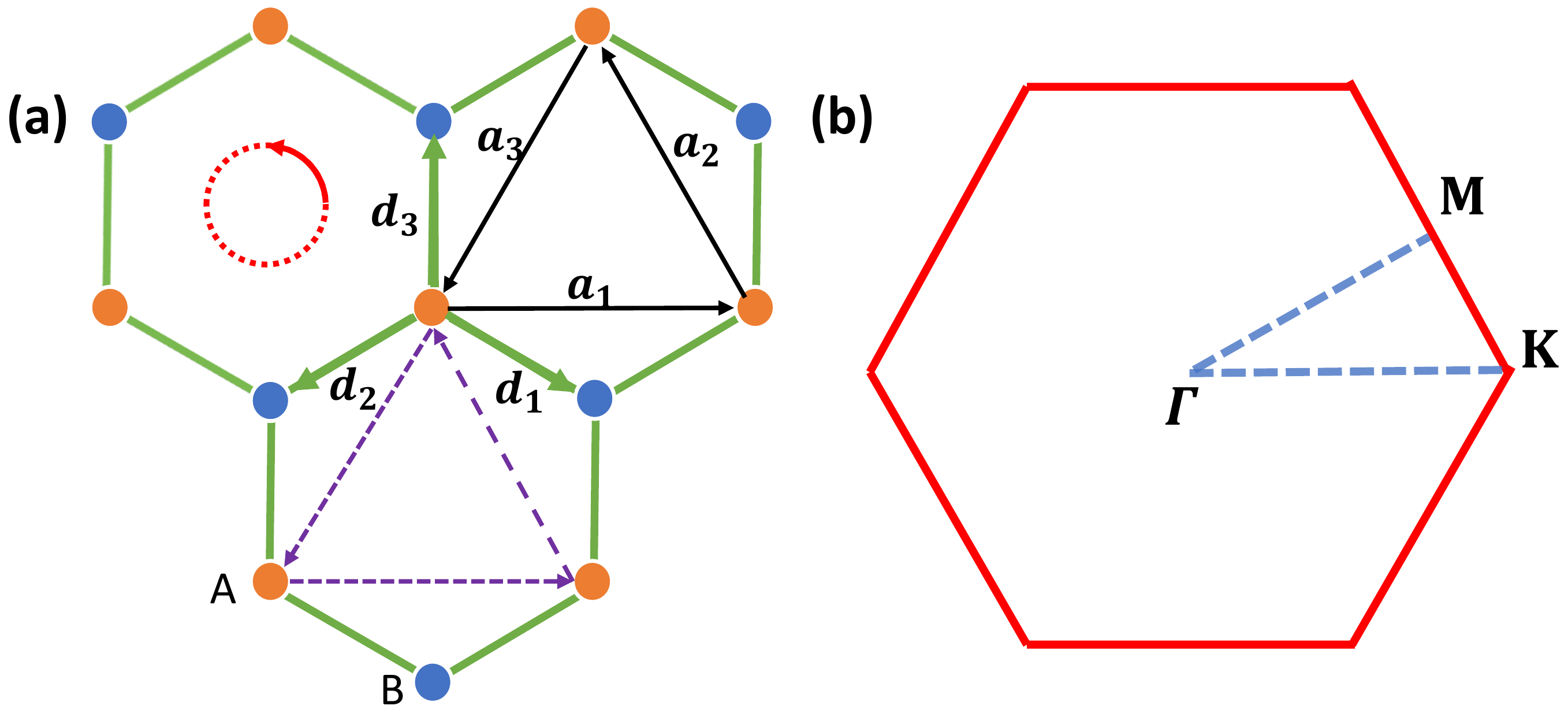}
\caption{(a) An AFM honeycomb with DMI.  The lattice vectors are $\bm a_1$, $\bm a_2$, and $\bm a_3$, and the nearest bond vectors are $\bm d_1$, $\bm d_2$, and $\bm d_3$.  (b) The corresponding hexagonal Brillouin zone.}
\label{fig:illustration}
\end{figure}  
 
\subsection{Honeycomb AFM}

\label{spin_Hamiltonian}

We begin with the following spin Hamiltonian on a honeycomb lattice,
\begin{equation} \label{eq:Hamiltonian}
\begin{split}
H &= J_{1}\sum_{\langle i,j\rangle}\bm{S_i}\cdot\bm{S_j} + D_2 \sum_{\langle\langle i,j\rangle\rangle} v_{ij} \hat{\bm z} \cdot(\bm{S_i}\times\bm{S_j}) \\
&\quad - h_{st}\sum_i (-1)^{i} S_{i}^{z} \;.
\end{split}
\end{equation}  
The first term describes the antiferromagnetic nearest-neighbor (NN) Heisenberg exchange with $J_{1}>0$. The second term is a second-NN DMI.  Here $v_{ij}=2\sqrt{3}(\bm d_1 \times \bm d_2)_z = \pm 1$ with $\bm d_1$ and $\bm d_2$ the vectors connecting site $i$ to its second NN site $j$, as shown in Fig.~\ref{fig:illustration}.  This second-NN DMI is allowed by crystal symmetry~\cite{Dzyaloshinsky1958, moriya1960}; it is intrinsic to the honeycomb lattice.  The third term is a staggered Zeeman field along the $z$ direction that stabilizes the system in the collinear AFM ground state at low temperatures~\footnote{While it is not easy to apply such a field externally, similar effects could arise when the SU(2) symmetry is broken by the single-ion anisotropy for $S>1/2$ or the Ising-type anisotropy in the exchange coupling within a Schwinger boson (SB) mean-field approach~\cite{timm2000}. This allows magnetic ordering in low-dimensional systems at finite temperature.}. Throughout this paper, we will use $J_1$ as the energy and temperature unit.

In the high-temperature paramagnetic (PM) phase, the low-energy spin dynamics can be described by spinons. We introduce the Schwinger boson (SB) representation for the spin operator~\cite{auerbach209}
\begin{equation}
\bm{S}_{i}\equiv\frac{1}{2}\sum_{s,s'}c^{\dagger}_{i,s}\bm{\sigma}_{ss'}c_{i,s'},
\quad (s, s' = \pm 1),
\label{eq:spinons}
\end{equation}
with the constraint that the number of spinons must be conserved at any given site, $\sum_{s}c^{\dagger}_{i,s}c_{i,s}=2S$. The index $s=\pm 1$ denotes up or down spins. In Eq.~\eqref{eq:spinons}$, \bm{\sigma}$ are the Pauli matrices, and $c^{\dagger}_{i,s}$ ($c_{i,s}$) denotes the creation (annihilation) operator for a spinon with spin $s$ at site $i$.  The spin amplitude $S=1/2$ is considered in this paper. 

Substituting Eq.~\eqref{eq:spinons} into the spin Hamiltonian~\eqref{eq:Hamiltonian}, we obtain
\begin{equation} \label{eq:Hsb}
\begin{split}
H_\text{SB}&=-2J_{1}\sum_{\langle i,j\rangle}\overrightarrow{\mathcal{A}}_{ij}^{\dagger}\overrightarrow{\mathcal{A}}_{ij}-\frac{i D_{2}}{2}\sum_{\langle\langle i,j\rangle\rangle}\sum_{s}sv_{ij}\mathcal{F}_{ij,s}^{\dagger}\mathcal{F}_{ij,-s} \\
&-h_{st}\sum_{is}\frac{(-1)^i}{2}sc_{i,s}^{\dagger}c_{i,s}+\sum_{i}\mu_{i}\Bigl (\sum_{s}c_{i,s}^{\dagger}c_{i,s}-2S\Bigr),
\end{split}
\end{equation}
where $\overrightarrow{\mathcal{A}}_{ij}\equiv (c_{i,\uparrow}c_{j,\downarrow}-c_{i,\downarrow}c_{j,\uparrow})/2$ is the antiferromagnetic NN bond operator, and $\mathcal{F}_{ij,s}\equiv c^{\dagger}_{is}c_{js}$ is the second NN bond operator. $\mu_i$ is a Lagrange multiplier to impose the local constraint at the mean-field level.
We note that $\overrightarrow{\mathcal{A}}_{ij} = -\overrightarrow{\mathcal{A}}_{ji}$ is antisymmetric.  Next we perform the mean-field decomposition of the quartic terms of the spinon Hamiltonian.  For the NN bond operator, we set $\bracket{\overrightarrow{\mathcal{A}}_{ij}}=- \bracket{\overrightarrow{\mathcal{A}}_{ji}} = \chi_{ij}$.  While, in general, $\chi_{ij}$ is complex, we work in the gauge in which $\chi_{ij}$ is real.  The second-NN order parameter can be written as $\langle\mathcal{F}_{ij,s}\rangle\equiv \eta^{S}_{ij,s}+iv_{ij}\eta^{A}_{ij,s}=\eta_{ij,s}$, where $\eta_{ij,s}^S = \bracket{\mathcal F_{ij,s} + \mathcal F_{ji,s}}/2$, and $\eta_{ij,s}^A = v_{ij}\bracket{\mathcal F_{ij,s} - \mathcal F_{ji,s}}/(2i)$.  The resulting bosonic Bogoliubov–de Gennes (BdG) Hamiltonian is given by
\begin{equation}  \label{Real-SB-Hamiltonian}
\begin{split}
H_{SB}^{M} & =-J_{1}\sum_{\langle i,j\rangle}\sum_{s}\Bigl(s\chi_{ij}c_{i,s}^{\dagger}c_{j,-s}^{\dagger}+\rm{H.c.}\Bigr) \\
 & +D_{2}\sum_{\langle\langle i,j\rangle\rangle}\sum_{s}\frac{i v_{ij}}{2}s\eta_{ij,-s}^{S}\Bigl(c_{i,s}^{\dagger}c_{j,s}-\rm{H.c.}\Bigr) \\
 & +D_{2}\sum_{\langle\langle i,j\rangle\rangle}\sum_{s}\frac{s}{2}\eta_{ij,-s}^{A}\Bigl(c_{i,s}^{\dagger}c_{j,s}+\rm{H.c.}\Bigr) \\
 &+\sum_{is}(\mu_i-\frac{(-1)^i h_{st}}{2}s)c_{i,s}^{\dagger}c_{i,s}, 
\end{split}
\end{equation}
where the trivial constant terms such as $2J_1 \sum_{\langle i,j \rangle} \chi_{ij}^2$ are neglected for simplicity.

This Hamiltonian can be simplified by symmetry considerations.  The spin Hamiltonian~\eqref{eq:Hamiltonian} has the combined $\mathcal{T}\mathcal{I}$ symmetry, which persists even in the low-temperature AFM phase.  Therefore, it is natural to expect that the high-temperature PM phase also preserves the $\mathcal{T}\mathcal{I}$ symmetry.  For the purpose of symmetry analysis, it is convenient to introduce sublattice-specific notations.  We use $a_{i,s}$ and $b_{i,s}$ to denote the annihilation operators on the $A$ and $B$ sublattices, respectively.  The corresponding second-NN bond order parameter is then denoted by $A_{ij,s}$ and $B_{ij,s}$.  The $\mathcal T$ and $\mathcal I$ symmetry are defined as~\footnote{Note that our definition of the $\mathcal I$ operator has an additional matrix $\sigma_3$.  It flips the sign of the spinon operator on the $B$ site, and is needed to make sure the NN bond term ($\chi_{ij}$) transforms into itself under the $\mathcal{TI}$ operation.  The $\sigma_3$ matrix is allowed since there is an extra phase freedom in the spinon representation.} (more details in Appendix~\ref{appendix_symmetry})
\begin{align} \label{eq:symm}
&\mathcal{T}c_{i,s}\mathcal{T}^{-1}=i(\sigma_{2})_{s,s^\prime}c_{i,s^{\prime}},\\
&\mathcal{I}\begin{bmatrix} a_{i} \\ b_{i} \end{bmatrix} \mathcal{I}^{-1}
=\sigma_{3}\sigma_{1} \begin{bmatrix} a_{-i}\\ b_{-i} \end{bmatrix}. \label{eq:symm1}
\end{align}
Imposing the $\mathcal{TI}$ symmetry on the mean-field Hamiltonian~\eqref{Real-SB-Hamiltonian} yields
\begin{equation} \label{tmd}
A^{\ast}_{ij,-s}=B_{-i-j,s} \;.
\end{equation}

\label{SOP}
\begin{figure}[t]
\centering
\includegraphics[width=0.9\columnwidth]{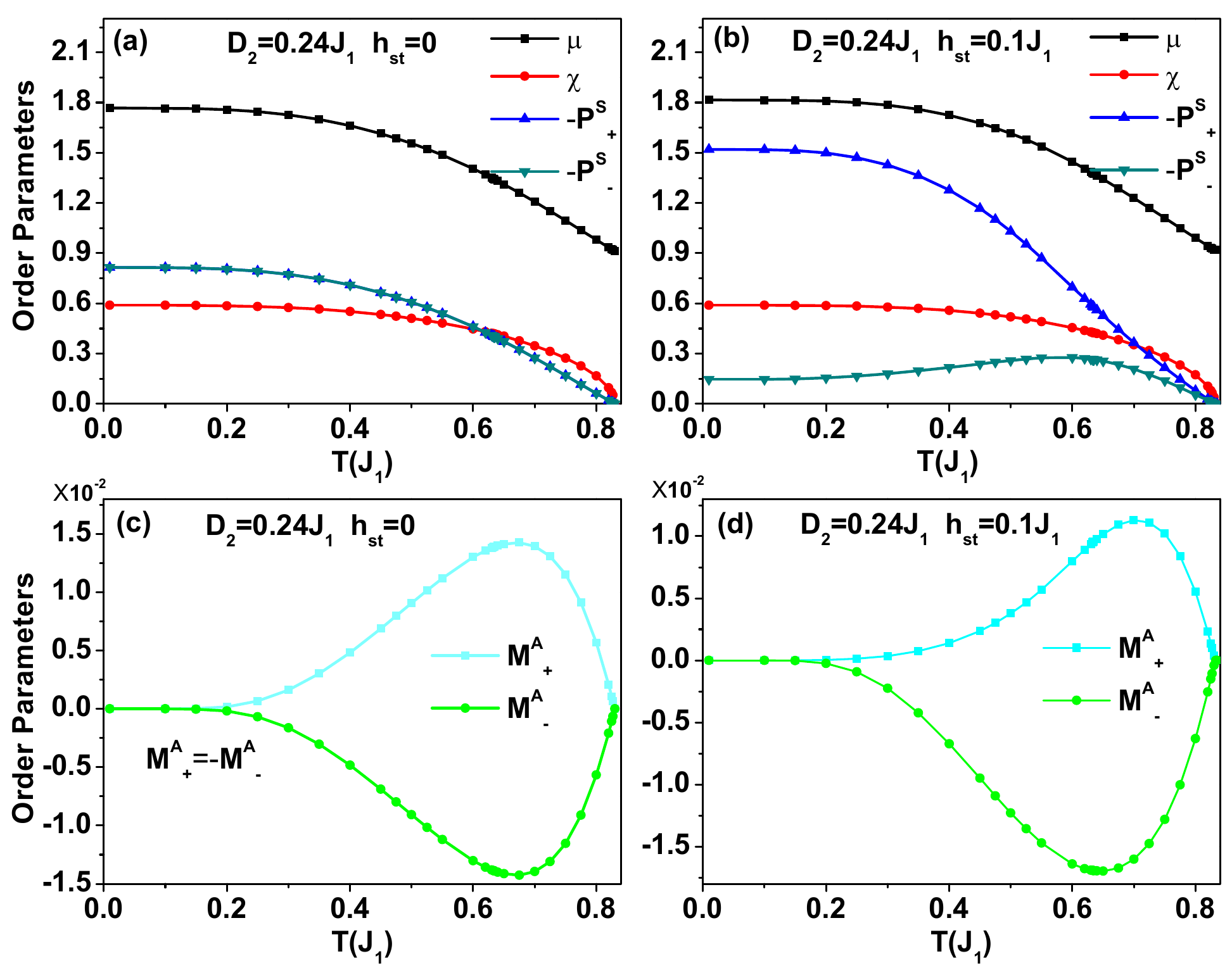}
\caption{The solution of order parameters with staggered field $h_{st}=0$ for (a) and (c), and $h_{st}=0.1J_{1}$ for (b) and (d).}
\label{fig:Ansatz}
\end{figure}

We now assume that the bond order parameters and the chemical potential are spatially uniform. They are $A_{ij,s}=A_{s}^{S}+iv_{ij}A^{A}_{s}$, $B_{ij,s}= B_{s}^{S}+iv_{ij}B^{A}_{s}$, $\chi_{ij}=\chi_{0}$, and $\mu_i=\mu$. Fourier transforming into the momentum space $\Psi_{\bm k s}=[a_{\bm k,s},b^{\dagger}_{-\bm k,-s}]^{T}=(1/\sqrt{N})\sum_{i}e^{-i\bm k\cdot \bm R_{i}}[a_{i,s},b^{\dagger}_{i,-s}]^{T}$, and using the condition~\eqref{tmd}, we obtain the mean-field spinon Hamiltonian in the momentum space
\begin{equation}
H^{M}_{SB}=\sum_{\bm k,s,\mu}\Psi^\dagger_{\bm k s}h^{s}_{\mu}(\bm k)\sigma_{\mu}\Psi_{\bm k s},
\label{eq:k-Hamiltonian}
\end{equation}
where $\sigma_{\mu}=\{I_{2\times 2}, \sigma_x, \sigma_y, \sigma_z\}$ and
\begin{subequations} \label{hhh}
\begin{align}
& h_{0}^{s}(\bm{k}) =\mu-s\frac{h_{st}}{2}+\frac{D_{2}s}{4} M_{-s}^{A}g_{S}(\bm{k}),\\
& h_{1}^{s}(\bm{k})-ih_{2}^{s}(\bm{k}) =-J_{1}\chi_{0}sf(\bm{k}),\\
& h_{3}^{s}(\bm{k}) =\frac{D_{2}s}{4}P_{-s}^{S}g_{A}(\bm{k}),
\end{align}
\end{subequations}
with $M_{s}^{A} \equiv A_{s}^{A} - B_{-s}^{A}$ and $P_{s}^{S} \equiv A_{s}^{S} + B_{-s}^{S}$.  The structure factors are $g_{A}(\bm k)\equiv-2\sum_{i}\sin(\bm k \cdot \bm a_{i})$, $g_{S}(\bm k)\equiv 2\sum_{i}\cos(\bm k \cdot \bm a_{i})$, and $f(\bm k)=\sum_{i}e^{i\bm d_{i}\cdot\bm k}$.  $g_A(\bm k)$ is an odd function of $\bm k$, and $g_S(\bm k)$ and $|f(\bm k)|$ are even functions of $\bm k$.

\begin{figure}[th!]
\centering
\includegraphics[width=1.0\columnwidth]{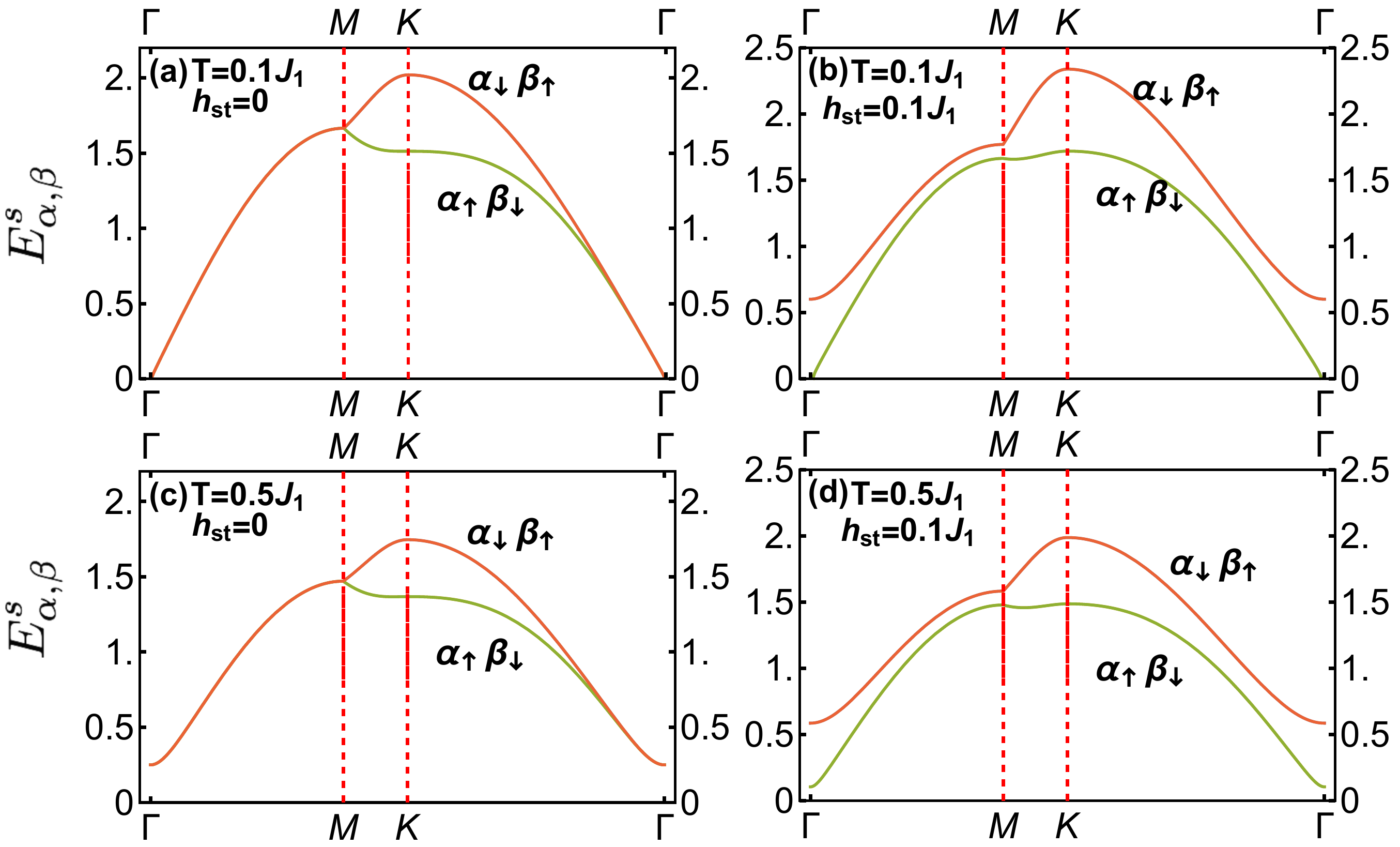}
\caption{The dispersions along high symmetry lines $\Gamma-M-K-\Gamma$ for (a) $h_{st}=0$ and $T=0.1J_{1}$, (b) $h_{st}=0.1J_{1}$ and $T=0.1J_{1}$, (c) $h_{st}=0$ and $T=0.5J_{1}$, and (d) $h_{st}=0.1J_{1}$ and $T=0.5J_{1}$. $\alpha(\beta)_{\uparrow (\downarrow)}$ denotes the mode $E^{s}_{\alpha(\beta)}(\bm k)$ with $s=\pm 1$ for spin $\uparrow (\downarrow)$.}
\label{fig:spectra}
\end{figure}

\subsection{Schwinger Boson mean-field solution}
The spinon Hamiltonian~\eqref{eq:k-Hamiltonian} contains six parameters that need to be determined self-consistently, namely, $\mu$, $\chi_0$, and $M_s^A$ and $P_s^S$ (with $s = \pm 1$).  
To diagonalize the Hamiltonian~\eqref{eq:k-Hamiltonian}, we perform the Bogoliubov transformation $\Phi_{\bm k, s}=U^{-1}_{s}(\bm k)\Psi_{\bm{k},s}=[\alpha_{\bm k, s},\beta^{\dagger}_{-\bm k,-s}]^{T}$, where $U^{-1}_s(\bm k)$ is a paraunitary matrix given by 
\begin{equation}
U^{-1}_{s}(\bm k)=\begin{bmatrix}
\cosh\frac{\theta_{s}(\bm k)}{2} & \sinh\frac{\theta_{s}(\bm k)}{2}e^{-i\varphi_{s}(\bm k)}\\
\sinh\frac{\theta_{s}(\bm k)}{2}e^{i\varphi_{s}(\bm k)} & \cosh\frac{\theta_{s}(\bm k)}{2} \end{bmatrix} \;.
\label{Umatrix}
\end{equation}
Here the Bogoliubov angles $\theta$ and $\varphi$ are defined by $h^s$ in Eq.~\eqref{hhh}: $h^{s}_{1}=h^{s}\sinh\theta_{s}\cos\varphi_{s}$, $h^{s}_{2}=h^{s}\sinh\theta_{s}\sin\varphi_{s}$ and $h^{s}_{0}=h^{s}\cosh\theta_{s}$, with $h^{s}\equiv\sqrt{h^{s2}_{0}-h^{s2}_{1}-h^{s2}_{2}}$.  The diagonalized Hamiltonian has the form $H^{M}_{SB}=\sum_{\bm k s}(E^{s}_{\alpha}(\bm k)\alpha^{\dagger}_{\bm k s}\alpha_{\bm k s}+E^{s}_{\beta}(\bm k)\beta^{\dagger}_{\bm k s}\beta_{\bm k s})$.  It is clear that $H^M_{SB}$ has two degenerate modes with $E^{s}_{\alpha}(\bm k)=E^{-s}_{\beta}(\bm k)=h^{s}(\bm k)+h^{s}_{3}(\bm k)$,
\begin{equation}
\begin{split}
E_{\alpha}^{s}(\bm{k}) &=\frac{D_{2}s}{4}P_{- s}^{S}g_{A\bm k} \\
&\quad + \sqrt{(\mu - s \frac{h_{st}}{2}+\frac{D_{2}s}{4}M_{- s}^{A}g_{S\bm k})^{2}-|J_{1}\chi_{0}f_{\bm k}|^{2}} \;.
\label{spinon-spectra}
\end{split}
\end{equation}
The wave function of the  $\alpha_{\bm k s}$ ($\beta_{\bm k s}$) quasiparticle is given in Appendix~\ref{BdG}.

This degeneracy originates from the combined $\mathcal{T}\mathcal{I}$ symmetry of our mean-field Hamiltonian.  We note that the annihilation operator of a spinon $\alpha_{\bm k s}$ transforms into into $s\beta_{\bm k,-s}$ under the $\mathcal{TI}$ operation defined in Eq.~\eqref{eq:symm}.
From this, we find
\begin{equation} \label{qq}
E^{s}_{\alpha}(\bm k)=E^{-s}_{\beta}(\bm k) \;.
\end{equation}
We call such a pair of degenerate modes as a $\mathcal{T}\mathcal{I}$ symmetry conjugate pair. This conjugate pair is crucial for the appearance of a pure transverse spin current as we discuss below.

We compute mean-field order parameters by solving a set of self-consistent equations detailed in Appendix~\ref{appendix_Self}.  The temperature dependence of order parameters at $D_{2}=0.24 J_1$ with different $h_{st}$ are shown in Fig.~\ref{fig:Ansatz}, along with the spinon dispersion in Fig.~\ref{fig:spectra}.  We first note that all order parameters vanish above $T_c \sim 0.826J_{1}$.  This is an artifact of the mean-field approach, and $T_c$ should be interpreted as a characteristic crossover temperature above which the system behaves as a paramagnet with local moments~\cite{auerbach209}.  On the other hand, as the temperature approaches zero, the spinon gap at the $\Gamma$ point closes (Fig.~\ref{fig:spectra}), and the system undergoes a phase transition into the collinear AFM phase at the N\'eel temperature $T_N$ via the spinon condensation~\cite{sarker1989}.  

For the current two-dimensional model, $T_N$ is strictly zero because single-site spin anisotropy or anisotropic exchange coupling is absent. Spin ordering at finite $T$ is mimicked by the nonzero staggered field $h_{st}$.

\section{Spin Nernst effect of spinons} 
\label{Top_Spinons}

\subsection{Spin conservation and mirror symmetry}

With a firm understanding of the spinon spectra, we now turn to the SNE. As a first step, we examine how many spins are carried by the spinon modes.  In general, this is not a trivial question, because in the presence of the DMI the spin angular momentum does not have to be conserved.  Fortunately, our model also has the mirror symmetry $\mathcal{M}_{z}$ about the lattice plane, which leads to the conservation of the total spin $S_z$,
\begin{equation}
S_{z}=\frac{\hbar}{2}\sum_{\bm k s}s\Psi^{\dagger}_{\bm k s}\sigma_{z}\Psi_{\bm k s}=\frac{\hbar}{2}\sum_{\bm k s}s\Phi^{\dagger}_{\bm k s}\sigma_{z}\Phi_{\bm k s}.
\end{equation}
We see that the $\alpha_{\bm k s}$ and $\beta_{\bm k -s}$ modes have opposite angular momentum $\langle 0|\alpha_{\bm k s}S_{z}\alpha^{\dagger}_{\bm k s}|0\rangle=\hbar s/2$ and $\langle 0|\beta_{\bm k -s}S_{z}\beta^{\dagger}_{\bm k -s}|0\rangle=-\hbar s/2$, respectively. Here $|0\rangle$ is the vacuum state of spinons.  The SNE is due to the opposite transverse motion of the two spin species driven by a longitudinal temperature gradient.

\subsection{Spin Nernst Effect coefficient in disordered state}
\label{SNE}
Since spinons do not carry charge, they cannot be driven by an external electric field, but they can respond to a statistical force, such as the temperature gradient $\bm\nabla T$. Due to the conservation of $S_z$, spin current can be written as $\bm{J}^{SN}=\sum_{s,\lambda}s(\hbar/2)\bm{J}^{s}_{\lambda}$, where $\bm J_\lambda^s$ is the spinon current of mode $\lambda$ and spin $s$.  According to the authors of Refs.~\cite{matsumoto2011a,matsumoto2011b,Lee2015,cheng2016}, the transverse $\bm J_\lambda^s$ due to $\bm\nabla T$ is given by
\begin{equation}
\label{species}
\bm{J}^{s}_{\lambda}=\frac{\hat{\bm z}}{\hbar}\times\nabla T\int \frac{d\bm{k}}{(2\pi)^2} c_{1}(n^{\lambda}_{s}(\bm k))\Omega^{s}_{\lambda}(\bm k) \;,
\end{equation}    
where $c_1$ is the weight function $c_{1}(x)=x\ln x-(1+x)\ln(1+x)$, and $n^{\lambda}_{s}(\bm k)$ and $\Omega^{s}_{\lambda}(\bm k)$ are the Bose-Einstein distribution function and the Berry curvature (defined below) for the mode $E^{s}_{\lambda}(\bm k)$, respectively.

\begin{figure}[tb]
\centering
\includegraphics[width=1.0\columnwidth]{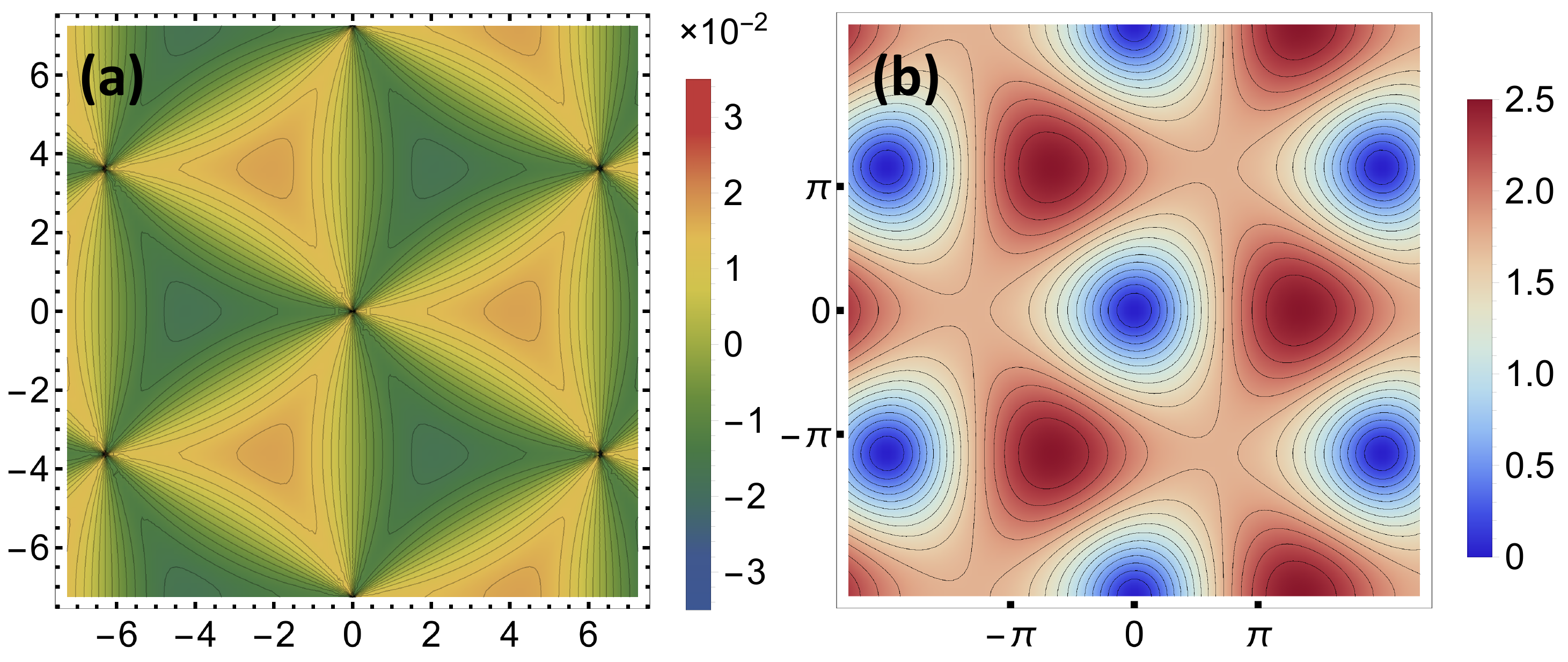}
\caption{The distributions of Berry curvature and spectrum for $\alpha_{\bm k, s}$ spinon with spin $s=-1$ at temperature $T=0.1J_{1}$ without staggered fields: (a) The Berry curvature; (b) The spectrum.}
\label{fig:berry}
\end{figure}

We now analyze the symmetry properties of the Berry curvature, which for the mode $E_{\alpha}^{s}(\bm k)$ is expressed as 
\begin{equation}
\begin{split}
\Omega^{s}_{\alpha}(\bm k)&=i\partial_{\bm{k}}u_{\alpha}^{s\dagger}(\bm{k})\times\sigma_{3}\partial_{\bm{k}}u^{s}_{\alpha}(\bm{k}) \\
&=\frac{1}{2}\nabla_{\bm k} \cosh \theta_{s}(\bm k)\times\nabla_{\bm k}\varphi_{s}(\bm k),
\label{bc}
\end{split}
\end{equation}
where $u^{s}_{\alpha}(\bm k)$ is the wave function of the $\alpha_{\bm k, s}$ quasiparticle as presented in Appendix~\ref{BdG}.  
Under the $\mathcal{TI}$ operation, $\alpha \to \beta$, $s\to -s$, and $\bm k \to \bm k$.  In addition, the Berry curvature should also flip sign due to the factor $i$ in its definition.  As such, under the $\mathcal{TI}$ operation, we have
\begin{equation} \label{xxx}
\Omega^{-s}_{\beta}(\bm k)=-\Omega^{s}_{\alpha}(\bm k) \;. 
\end{equation}
Together with the energy dispersion relation $E^{s}_{\alpha}(\bm k)=E^{-s}_{\beta}(\bm k)$ [see Eq.~\eqref{qq}], this relation indicates that $\bm J_\alpha^s$ and $\bm J_\beta^{-s}$ are always opposite in sign, resulting in a pure transverse spin current.

Next we focus on a particular mode $\alpha$.  For bosonic BdG equations, there is a general relation of the Berry curvature between the $\alpha$ and $\beta$ mode (see Appendix~\ref{appendix_BC})
\begin{equation}
\Omega_\beta^s(\bm k) = \Omega_\alpha^{-s}(-\bm k) \;.
\end{equation}
Combining this relation with Eq.~\eqref{xxx}, we have
\begin{equation}
\Omega^{s}_{\alpha}(\bm k)=-\Omega^{s}_{\alpha}(-\bm k) \;.
\end{equation}
This is clearly seen in Fig.~\ref{fig:berry} (a). If the spinon dispersion is inversion symmetric, the transverse current $\bm J_\alpha^s$ would vanish. However, as we can see from Eq.~\eqref{spinon-spectra}, the presence of the DMI breaks this symmetry, i.e., $E^{s}_{\lambda}(\bm k)\neq E^{s}_{\lambda}(-\bm k)$ as illustrated in Fig.~\ref{fig:berry} (b). After summing over all occupied states, there should be a net spinon current.  Therefore the second NN DMI is crucial for the appearance of the SNE.
   
\begin{figure}[tb]
\centering
\includegraphics[width=0.49\columnwidth]{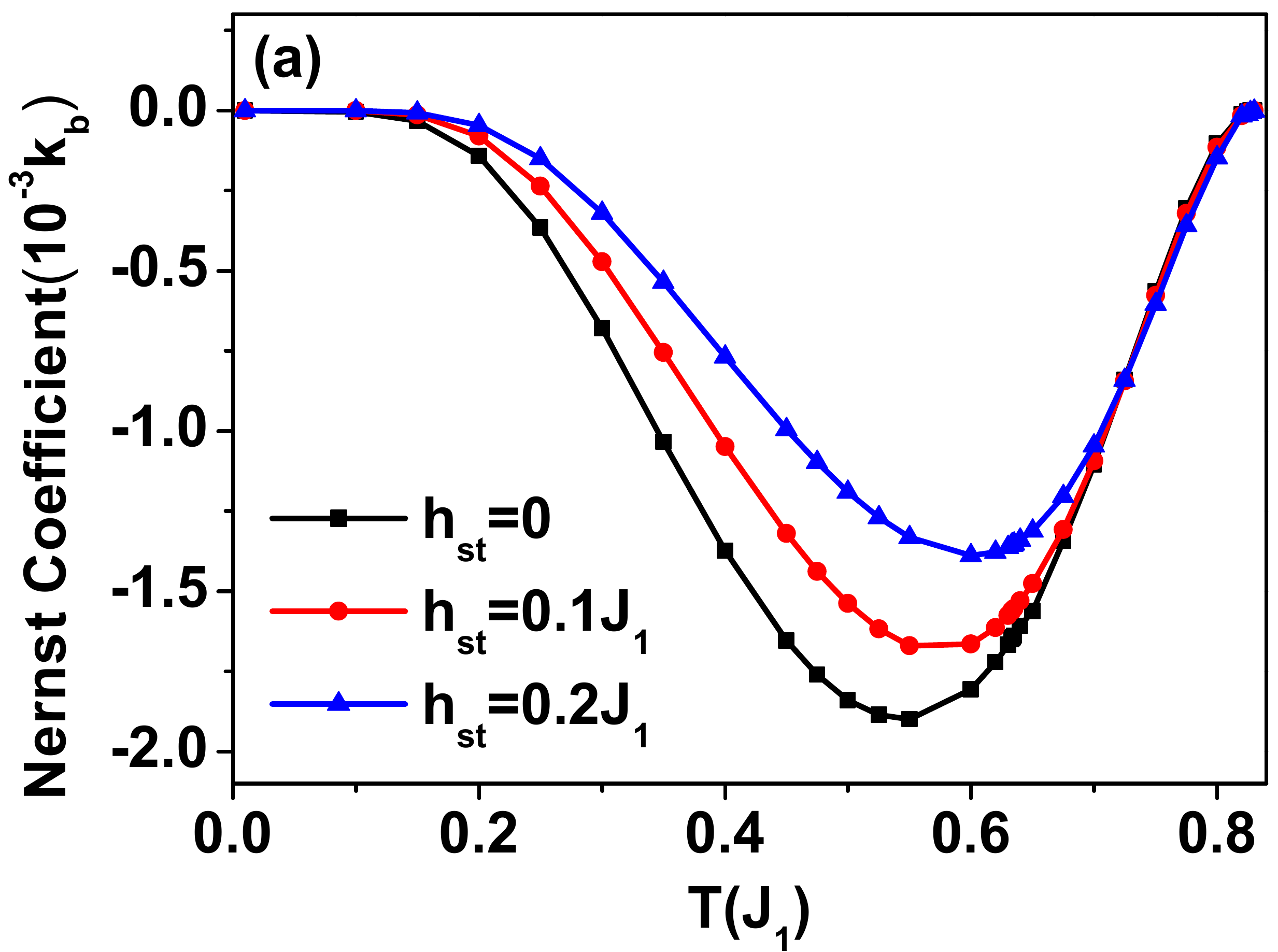}
\includegraphics[width=0.49\columnwidth]{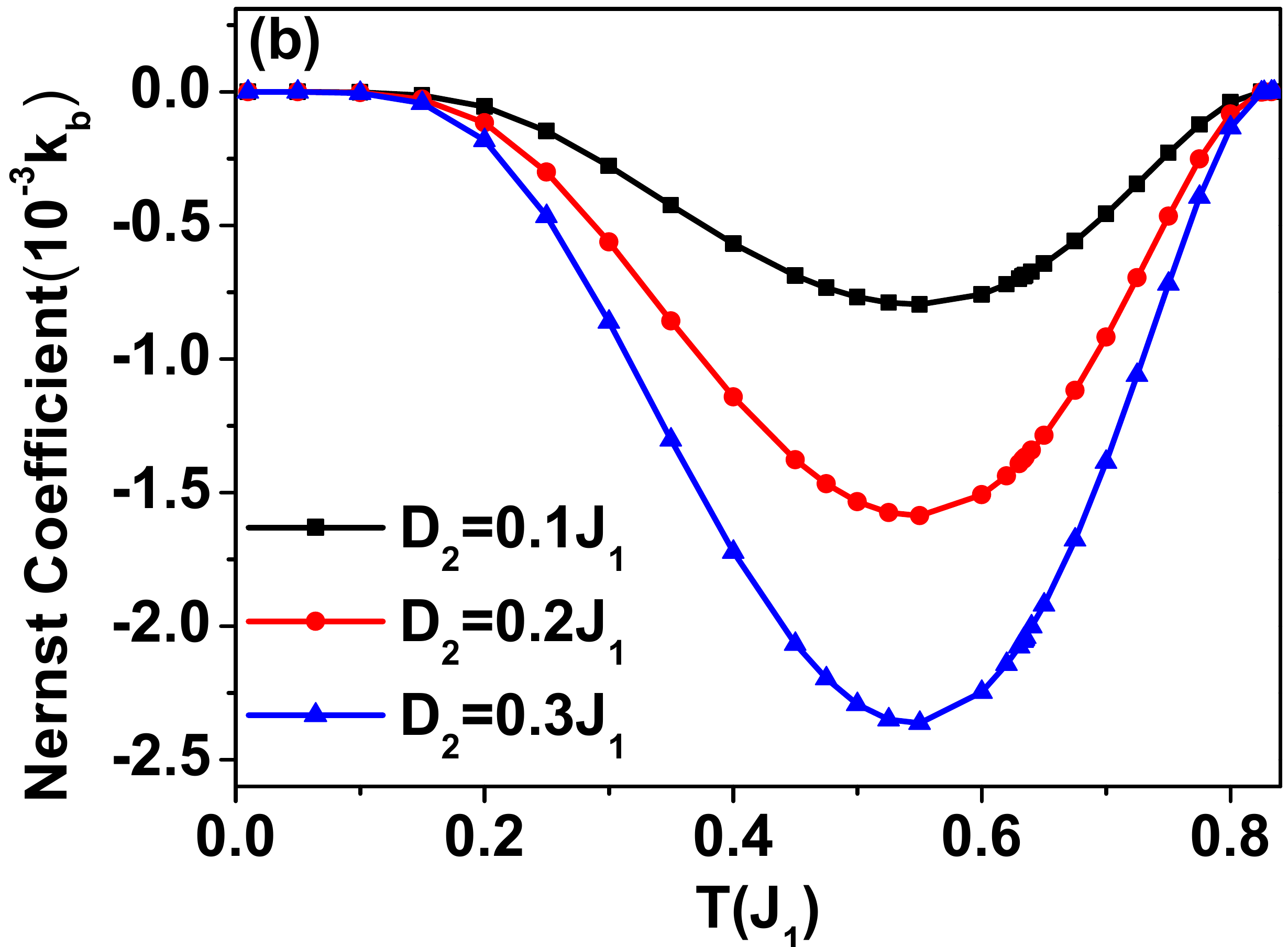}
\caption{The spinon Nernst coefficients as a function of temperature: (a) at different staggered fields $h_{st}$ and a fixed DM interaction $D_{2}=0.24J_{1}$;and (b) at different DM interaction $D_2$ and without a staggered field.}
\label{fig:NC}
\end{figure} 
    
We numerically calculate the spin Nernst coefficient given by~\cite{matsumoto2011a,matsumoto2011b,Lee2015,cheng2016}
\begin{equation}
\alpha_{xy}=\sum_{s}\int \frac{d\bm{k}}{(2\pi)^2} c_{1}(n^{\alpha}_{s}(\bm k))\Omega^{s}_{\alpha}(\bm k).
\label{SNE_C}
\end{equation}
where $\alpha_{xy}$ is defined by the relation $\bm{J}^{SN}=\alpha_{xy}\hat{\bm{z}}\times\bm\nabla{T}$. The temperature dependence of $\alpha_{xy}$ is calculated at different staggered field $h_{st}$ and DMI strength $D_{2}$ in Fig.~\ref{fig:NC}. 
We find that $\alpha_{xy}$ will be zero at two ends of the temperature zone, i.e., $T=0$ and $T=T_{c}$. 
When $T$ approaches zero, the fluctuating component of spinons is decreased. On the other hand, when the temperature approaches $T=T_{c}$, $P_s$ is reduced to zero.  This will cause the SNE to vanish because the vanishing of $P_{s}$ effectively restores the inversion symmetry of the spinon dispersion.

In addition, the peak of the spin Nernst coefficient at a special temperature results from the competition between the enhancement of excited spinons engaging in transport and the reduction of the second-NN order parameter $P_{s}$ and $M_s$ as the temperature increases. 
The staggered field will weaken the spin Nernst coefficient in opposite to that of DMI, because the staggered field supports a collinear configuration, but DMI favors a perpendicular one between two second-NN spin polarizations. In reality, $T_N$ could be finite due to a variety of effects neglected here, 
and the temperature dependence of the spin Nernst coefficient is expected to depend on the competition between these effects and the DMI, especially near $T_N$. Nevertheless, the spin Nernst coefficient is shown to change continuously with increasing $h_{st}$. This implies that the spin Nernst coefficient changes continuously at the magnetic transition temperature as long as it is the second-order transition.

\subsection{The relation between magnons and spinons in antiferromagnets}

\label{CON}

We now explore the connection between the SNE in the PM phase and the SNE in the AFM phase.  In the Schwinger boson picture, the transition from the PM to AFM phase takes place via the spinon condensation~\cite{sarker1989}.  Take Fig.~\ref{fig:spectra}(b) as an example.  As the temperature is lowered to the N\'eel temperature $T_N$, the spinons will condense into the $\alpha_\uparrow$ and $\beta_\downarrow$ modes.  Consequently, the resulting state will have a macroscopic occupation of spin up (down) at A (B) sites, giving rise to the AFM order.  At the same time, the two upper modes, $\alpha_\downarrow$ and $\beta_\uparrow$, will evolve into magnons.  In fact, upon the spinon condensation, the order parameter $M_{-s}^A$ vanishes, and the dispersion of the $\alpha_\downarrow$ mode becomes
\begin{equation}
E_{\alpha\downarrow}(\bm k) =-\frac{D_{2}}{4}P_{+}^{S}g_{A}(\bm{k})+\sqrt{(\mu+ h_{st}/2)^2-|J_{1}\chi_{0}f(\bm{k})|^{2}} \;.
\end{equation}
Comparing the above expression with that of magnons $E_{m}(\bm{k})=SD_{2}g_{A}(\bm{k}) + \sqrt{(J_{1}S+h_{st})^{2}-S^{2}J_{1}^{2}|f(\bm{k})|^{2}}$~\cite{cheng2016,zyuzin2016}, we see that they share the basic algebraic structure.  The slight difference is due to the incomplete condensation of spinons.  

\begin{figure}[t]
\centering
\includegraphics[width=1.0\columnwidth]{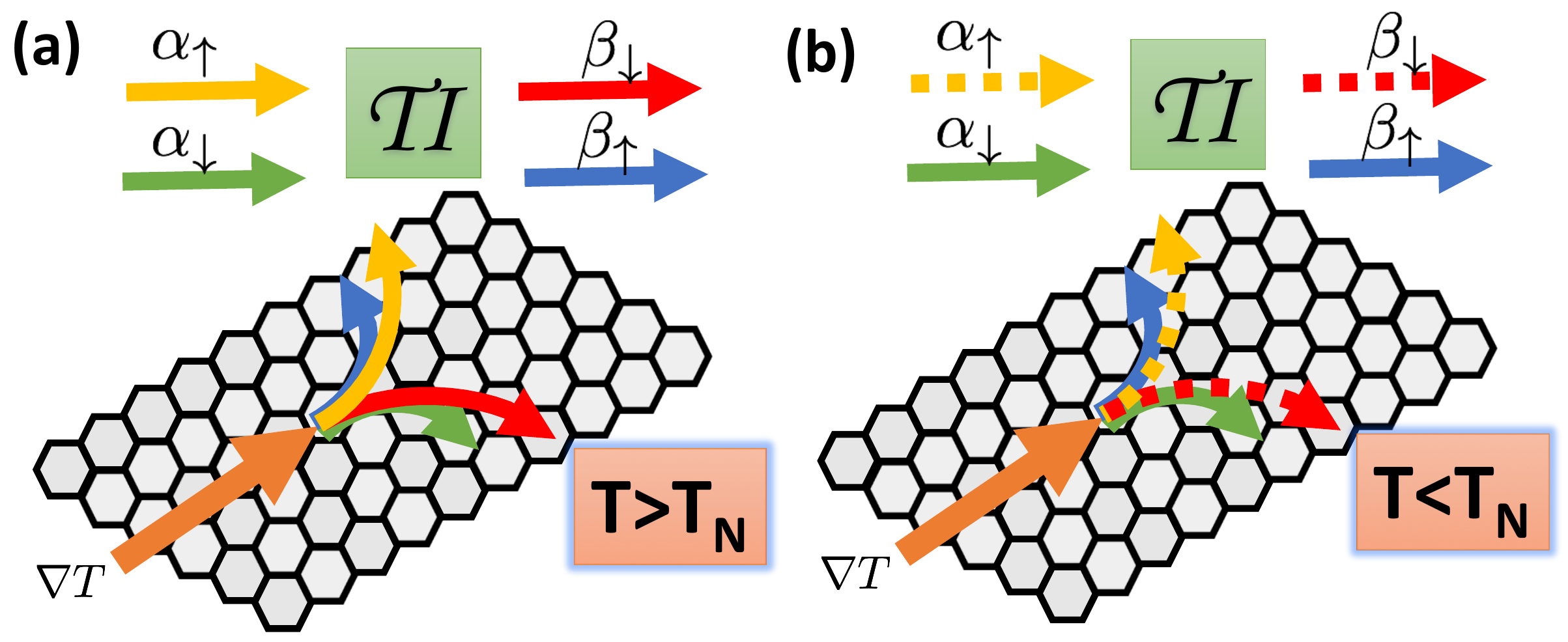}
\caption{It illustrates the spin Nernst effect on a honeycomb AFM carried by spin fluctuations: (a) at $T>T_{N}$, the SNE is carried by spinons in the paramagnetic phase, and (b) at $T<T_{N}$, the SNE is carried by magnons.}
\label{fig:Transport}
\end{figure}

It is obvious that across the phase boundary between the PM and AFM phase, the symmetries relevant to the SNE, namely, the combined $\mathcal{TI}$ symmetry and the breaking of the spin rotational symmetry due to $D_2$ remains the same.  Hence the SNE in both the PM and AFM phase has the same microscopic origin, as shown in Fig.~\ref{fig:Transport}. 

\section{Summary and discussion}
\label{Disc}
In summary, we study the pure SNE in the PM state on an antiferromagnetic honeycomb lattice with a second-NN DMI, using the Schwinger boson mean-field method. We find that the pairs of the combined $\mathcal{T}\mathcal{I}$ conjugate modes of spinons support a transverse spin current without a transverse thermal current. Because of the competition between the short-range spin correlations, represented by the temperature-dependent mean-field order parameters, and 
spin fluctuations, represented by the thermal population of spinons, the spin Nernst coefficient shows a nontrivial temperature dependence for a rather simple model considered here.
This might suggest that a paramagnetic insulator with AFM interaction of spins could serve as a spintronics device even above the magnetic transition temperature to generate or detect the spin current.  

Before closing, we would like to discuss several issues left for future studies. Throughout this paper, we neglect the fluctuations from the mean-field solution. In fact, the Schwinger boson mean-field treatment is the result of the zeroth order of $O(1/N)$ in a large-$N$ expansion of a spin SU($N$) model~\cite{auerbach209}. Rigorously speaking, the low energy part of fluctuations, i.e., the phase fluctuation of order parameters, could couple with the $U(1)$ gauge field, the dynamics of which may exhibit a confined or deconfined phase. Exploring these effects of fluctuations on spin transport will be an interesting problem in the future~\cite{read1991,zhou2017}. However, since our argument about the finite SNE in the PM state of the honeycomb AFM is based on the combined $\cal T \cal I$ symmetry, our conclusion would not be altered in a qualitative manner.

We do not use the full projected symmetry group method to analyze the spinon Hamiltonian. 
Such analyses would be necessary for spin liquid systems at low temperatures described by fermionic spinons.
On the other hand, for investigating the pure SNE at high temperatures, 
it is sufficient to consider only the combined $\mathcal{T}\mathcal{I}$ symmetry based on the unprojected spinon wave function.

So far we only considered the so-called intrinsic contribution to the spinon SNE due to the Berry curvature of the spinon bands.  Similar to the anomalous Hall effect~\cite{nagaosa2010}, there should be extrinsic effect due to the scattering between spinons and other relevant physical degrees of freedom such as phonons.  We note that there is an analogous effect of electrons~\cite{gusynin2014,gusynin2015} for which the impurity scattering has been discussed~\cite{dyrda2016}.

In real materials, such as transition-metal trichalcogenides, the situation is more complicated. 
In addition to the interactions described in Eq.~(\ref{eq:Hamiltonian}), longer-range exchange interactions are present, stabilizing complex magnetic ordered states~\cite{sivadas2015}. 
Furthermore, single-ion anisotropies and anisotropic exchange interactions could exist, making finite-temperature magnetic ordering possible even for the two-dimensional limit~\cite{huang2017}. 
These effects not only require solving a set of self-consistent equations for many order parameters, 
but also require extending the current formalism as demonstrated in Ref.~\cite{timm2000}. 
For $S>1/2$ systems, $h_{st}$ is related to the single-ion anisotropy $K_2$ as $h_{st}\sim K_2(S-1/2)/S M_{z}$ with 
$M_z=\sum_{s,s^\prime} (\sigma_3)_{s,s^\prime} \langle c_{i, s}^\dag c_{i, s^\prime} \rangle$~\cite{timm2000}.
For $\rm{MnPS_3}$ as discussed in Ref.~\cite{cheng2016}, $h_{st} /J$ could become as large as $~0.01$ at low temperatures. This value is an order of magnitude smaller than the ones used in our analyses. Therefore, it is expected that the spin Nernst coefficient does not change significantly across a magnetic transition temperature. 
Detailed material dependence of the SNE including these effects is left for future studies.

\acknowledgments{
We are grateful to Ying Ran for discussions. We also acknowledge useful discussions with Ran Cheng, Rina Takashima, Yang Gao, and Wenyu Shan.  Y.Z.\ and D.X.\ are supported by the Department of Energy, Basic Energy Sciences, Grant No.DE-SC0012509. S.O.\ acknowledges support by the U.S. Department of Energy, Office of Science, Basic Energy Sciences, Materials Sciences and Engineering Division.
}

\appendix

\section{Symmetry operations}
\label{appendix_symmetry}

We discuss symmetry operations on the spinon Hamiltonian in the momentum space.  These symmetry operations include inversion operation $\mathcal{I}$,  time reversal operation $\mathcal{T}$, and mirror operation $\mathcal{M}_{z}$. 

The spinon Hamiltonian matrix at each $\bm k$ point is given by 
\begin{equation}
H_{s}(\bm k)=\sum_{\mu} h^{s}_\mu(\bm k)\sigma_{\mu}.
\label{Hk_spinon}
\end{equation}
For the inversion operation, we follow the definition of Eq.~\eqref{eq:symm1} in the lattice space, which ensures that  
$\mathcal{I}\bm{S}_{i,A(B)}\mathcal{I}^{-1}=\bm{S}_{-i,B(A)}$. Accordingly, the Hamiltonian matrix $H_{s}(\bm k)$ is transformed as 
\begin{equation}
\mathcal{I}H_{s}(\bm{k})\mathcal{I}^{-1}=\sigma_{2}H_{-s}^{T}(\bm{k})\sigma_{2},
\end{equation}
where $T$ stands for the matrix transposition.
 
The time reversal operator $\mathcal{T}$ is defined in Eq.~\eqref{eq:symm},
and transforms $H_{s}(\bm k)$ into
\begin{equation}
\mathcal{T}H_{s}(\bm k)\mathcal{T}^{-1}=\sigma_{3}H_{-s}^{\ast}(-\bm{k})\sigma_{3}.
\end{equation}
Under the combined $\mathcal{T}\mathcal{I}$ operation, $H_{s}(\bm k)$ is thus transformed as 
\begin{equation}
\mathcal{T}\mathcal{I}H_{s}(\bm{k})(\mathcal{T}\mathcal{I})^{-1}=\sigma_{1}H_s(-\bm k)\sigma_{1}.
\label{HTI}
\end{equation}
Therefore, if the system has the combined $\mathcal{T}\mathcal{I}$ symmetry, then $H_{s}(\bm k)$ should satisfy
\begin{equation}
\sigma_{1}H_s(-\bm k)\sigma_{1}=H_s(\bm k) \;.
\end{equation}

The mirror symmetry operator $\mathcal{M}_{z}$ with respect to the lattice plane is defined as 
\begin{equation}
\mathcal{M}_{z}c_{i,s}\mathcal{M}^{-1}_{z}=i(\sigma_{3})_{s,s^\prime}c_{i,s^{\prime}},
\end{equation}    
which leads to $\mathcal{M}_{z}S^{z}_{i}\mathcal{M}^{-1}_{z}=S^{z}_{i}$ and $\mathcal{M}_{z}S^{x,y}_{i}\mathcal{M}^{-1}_{z}=-S^{x,y}_{i}$. The Hamiltonian matrix is invariant under mirror operation $\mathcal M$
\begin{equation}
\mathcal{M}H_{s}(\bm k)\mathcal{M}^{-1}=H_{s}(\bm{k}) \;.
\end{equation}
  
\section{Mean-field self-consistent equations} 
\label{appendix_Self}
The mean-field order parameters and the Lagrange multiplier $\mu$ are determined by minimizing the free energy involving these parameters.
By differentiating the free energy with respect to these parameters and equating to zero, one arrives at the following set of self-consistent equations: 
\begin{subequations}
\begin{align}
\label{eq:self-consistent}
1+&2S =\frac{1}{2N}\sum_{\bm{k}s}\left[\frac{h_{0}^{s}(\bm{k})}{h^{s}(\bm{k})}(n^{\alpha}_{\bm{k},s}+n^{\beta}_{-\bm{k},-s}+1)\right],\\
4\chi_{0} & =\frac{J_{1}\chi_{0}}{3N}\sum_{\bm{k}s}\left[\frac{|f(\bm k)|^2}{h^{s}(\bm{k})}(n^{\alpha}_{\bm{k},s}+n^{\beta}_{-\bm{k},-s}+1)\right],\\
M_{s}^{A} & =\frac{1}{6N}\sum_{\bm{k}}g_{A}(\bm{k})(n^{\alpha}_{\bm{k},s}-n^{\beta}_{-\bm{k},-s}-1),\\
-P_{s}^{S} & =\frac{1}{6N}\sum_{\bm{k}}g_{S}(\bm{k})\frac{h_{0}^{s}(\bm{k})}{h^{s}(\bm{k})}(n^{\alpha}_{\bm{k},s}+n^{\beta}_{-\bm{k},-s}+1),
\end{align}
\end{subequations}
where $n^{\alpha/\beta}_{\bm k, s}=[\exp{(E^{s}_{\alpha,\beta}(\bm k)/T)}-1]^{-1}$ is the Bose distribution function, and $N$ is the number of unit cells. 

\section{BdG equation and Berry curvature}

\label{BdG}

In this section we present a detailed discussion of the bosonic BdG equation and the associated wave functions.  Our starting point is the spinon mean-field Hamiltonian~\eqref{eq:k-Hamiltonian}, reproduced here for convenience,
\begin{equation} \label{kkk}
H = \sum_{\bm k,s} \Psi_{\bm ks}^\dag H_s(\bm k) \Psi_{\bm ks} \;,
\end{equation}
where $\Psi_{\bm ks} = [a_{\bm k, s}, b_{-\bm k,-s}^\dag]^T$ with $a_{\bm k, s}$ and $b_{\bm k, s}$ being the Fourier transform of the spinon operators on the $A$ and $B$ sublattices, respectively.  Introduce the Bogoliubov transformation
\begin{equation} \label{bog}
\binom{a_{\bm k, s}}{b_{-\bm k,-s}^\dag} = U_s(\bm k) \binom{\alpha_{\bm k,s}}{\beta_{-\bm k, -s}^\dag}  \;.
\end{equation}
The boson commutation relation dictates that $U_s(\bm k)$ is a paraunitary matrix, i.e.,
\begin{equation}
U_s(\bm k) \sigma_3 U_s^\dag(\bm k) = \sigma_3 \;.
\end{equation}
By demanding that the Bogoliubov transformation diagonalizes the Hamiltonian, i.e., $H =\sum_{\bm ks}[E^s_\alpha(\bm k)\alpha^\dagger_{\bm ks}\alpha_{\bm ks}+E^s_\beta(\bm k)\beta^\dagger_{\bm ks}\beta_{\bm ks}]$, we obtain the BdG equation
\begin{equation}
H_{s}(\bm k)U_{s}(\bm k)=\sigma_{3}U_{s}(\bm k)\sigma_{3}\Delta(\bm k),
\label{initialeigen}
\end{equation}
where $\Delta(\bm k)=\mathrm{diag}(E^{s}_{\alpha}(\bm k), E^{-s}_{\beta}(-\bm k))$ is the eigenvalue matrix. We note that both the excitation energies $E^{s}_{\alpha}(\bm k)$ and $E^{-s}_{\beta}(-\bm k)$ must be positive. Otherwise the mean-field solution is unphysical.  The explicit expression of $U_s(\bm k)$ is given Eq.~\eqref{Umatrix}.

For the purpose of calculating the Berry curvature, it is necessary to clarify the wave function of a spinon quasiparticle.  Let us write $U_s(\bm k) = [u_\alpha^s(\bm k), u_{\bar\beta}^{-s}(-\bm k)]$, where $u_\alpha^s(\bm k)$ and $u_{\bar\beta}^{-s}(-\bm k)$ are two-component column vectors.  Inserting this expression into the BdG equation~\eqref{initialeigen}, we have
\begin{subequations} \label{qqq}
\begin{align}
H_s(\bm k) u_\alpha^s(\bm k) &= E_\alpha^s(\bm k) \sigma_3 u_\alpha^s(\bm k) \;, \\
H_s(\bm k) u_{\bar\beta}^{-s}(-\bm k) &= -E_\beta^{-s}(-\bm k) \sigma_3 u_{\bar\beta}^{-s}(-\bm k) \;.
\end{align}
\end{subequations}
It is clear that $u^s_\alpha(\bm k)$ and $u^{-s}_{\bar\beta}(-\bm k)$ are the wave functions of the quasiparticle $\alpha_{\bm k,s}$ with positive energy $E^s_{\alpha}(\bm k)$ and the quasihole $\beta_{-\bm k, -s}$ with negative energy $-E^{-s}_\beta(-\bm k)$, respectively.  We denote the quasihole wave functions by the subscript $\bar\alpha$ or $\bar\beta$.

The above discussion suggests that to find the quasiparticle wave function of the $\beta_{-\bm k, -s}$ mode, we just need to recast the spinon Hamiltonian in the basis $\tilde\Psi_{\bm ks} = [b_{\bm k, s}, a_{-\bm k,-s}^\dag]^T$.  To do that, we make use of the particle-hole conjugate operator, defined by
\begin{equation} \label{zzz}
\mathcal{C}c_{i,s}\mathcal{C}^{-1}=c^{\dagger}_{i,s}.
\end{equation}
Acting $\mathcal C$ on the basis $\Psi_{\bm ks}$, we have
\begin{equation}
\mathcal{C}\Psi_{\bm k s}\mathcal{C}^{-1}=\sigma_{1}\begin{bmatrix} b_{\bm k,-s}\\ a^{\dagger}_{-\bm k,s}\end{bmatrix} \;.
\end{equation}
Consequently,
\begin{equation}
\tilde H_s(\bm k) = \mathcal{C}H_{s}(\bm k)\mathcal{C}^{-1}=\sigma_{1}H^{\ast}_{s}(-\bm k)\sigma_{1}.
\label{Hpq}
\end{equation}
We can then deduce that 
\begin{equation}
u^{-s}_{\beta}(\bm k)=\sigma_{1}u^{-s\ast}_{\bar\beta}(\bm k) \;.
\label{wf-ph}
\end{equation}

If the system has $\mathcal{TI}$ symmetry, according to Eq.~\eqref{HTI}\begin{equation}
\tilde{H}_{s}(\bm k)=\sigma_{1}H^{\ast}_{s}(-\bm k)\sigma_1 = H^*_s(\bm k) \;.
\end{equation}
Since $\tilde H_s(\bm k)$ and $H^*_s(\bm k)$ describe the same physical system, we have
\begin{equation}
u^{-s}_{\beta}(\bm k)=u^{s\ast}_{\alpha}(\bm k) \;.
\label{wf-TI}
\end{equation} 

\section{The property of Berry Curvature}
\label{appendix_BC}
The Berry curvature is generally defined by the projection operator
\begin{equation}
\label{projectbc} 
\Omega_{n}(\bm k)=-i\epsilon^{ij}\mathrm{Tr}[\bar{P}_{n}(\bm k)\partial_{k_i}P_{n}(\bm k)\partial_{k_j}P_{n}(\bm k)] ,
\end{equation}
where $P_{n}(\bm k)$ is the projection operator for the $n$-th band at the momentum $\bm k$, and $\bar{P}_{n}\equiv 1-P_{n}$.  Note that for the generalized eigenvalue problem given by Eq.~\eqref{qqq}, the projector operator is defined by~\cite{shindou2013}
\begin{equation}
P_n = \frac{\ket{n}\bra{n}\sigma_3}{\bracket{n|\sigma_3|n}} \;.
\end{equation}

For our disordered AFM described by the bosonic BdG Hamiltonian $H_{s}(\bm k)$, this leads to the formula
\begin{equation}
\Omega^{s}_{\lambda}(\bm k)=i\partial_{\bm{k}}u_{\lambda}^{s\dagger}(\bm{k})\times\sigma_{3}\partial_{\bm{k}}u^{s}_{\lambda}(\bm{k})/(u_{\lambda}^{s\dagger}(\bm{k})\sigma_{3}u^{s}_{\lambda}(\bm{k})),
\end{equation} 
where $u_{\lambda,s}(\bm k)$ is the wave function of $\lambda$-type quasiparticle or quasihole, and the normalization $u_{\lambda,s}^{\dagger}(\bm{k})\sigma_{3}u_{\lambda,s}(\bm{k})=\pm 1$ for quasiparticle and quasiholes, respectively.

For a two-level system, it follows from Eq.~\eqref{projectbc} that the Berry curvature has the property 
\begin{equation}
\Omega_{n}(\bm k)=-\Omega_{\bar n}(\bm k),
\end{equation}
where $n$ and $\bar{n}$ refers to the quasiparticle and quasihole bands, respectively. This property is a special case of $\sum_{n}\Omega_{n}(\bm k)=0$ with $n\geq 2$. Applying this relation to our Hamiltonian $H_{s}(\bm k)$, we have 
\begin{equation}
\Omega^{s}_{\alpha}(\bm k)=-\Omega^{-s}_{\bar\beta}(-\bm k).
\end{equation}  

Using Eq.~\eqref{wf-ph}, one can deduce the relation
\begin{equation}
\Omega^{-s}_{\beta}(\bm k)=-\Omega^{-s}_{\bar\beta}(\bm k).
\label{bc-ph}
\end{equation} 
The result can be also applied to a general reduced BdG Hamiltonian.   
 
In the presence of the $\mathcal{T}\mathcal{I}$ symmetry, the Berry curvatures for the two modes $\alpha$ and $\beta$ could be also related. Using Eq.~\eqref{wf-TI}, we find
\begin{equation}
\Omega^{s}_{\alpha}(\bm k)=-\Omega^{-s}_{\beta}(\bm k).
\label{bc-TI}
\end{equation}




\begin{thebibliography}{44}%
\makeatletter
\providecommand \@ifxundefined [1]{%
 \@ifx{#1\undefined}
}%
\providecommand \@ifnum [1]{%
 \ifnum #1\expandafter \@firstoftwo
 \else \expandafter \@secondoftwo
 \fi
}%
\providecommand \@ifx [1]{%
 \ifx #1\expandafter \@firstoftwo
 \else \expandafter \@secondoftwo
 \fi
}%
\providecommand \natexlab [1]{#1}%
\providecommand \enquote  [1]{``#1''}%
\providecommand \bibnamefont  [1]{#1}%
\providecommand \bibfnamefont [1]{#1}%
\providecommand \citenamefont [1]{#1}%
\providecommand \href@noop [0]{\@secondoftwo}%
\providecommand \href [0]{\begingroup \@sanitize@url \@href}%
\providecommand \@href[1]{\@@startlink{#1}\@@href}%
\providecommand \@@href[1]{\endgroup#1\@@endlink}%
\providecommand \@sanitize@url [0]{\catcode `\\12\catcode `\$12\catcode
  `\&12\catcode `\#12\catcode `\^12\catcode `\_12\catcode `\%12\relax}%
\providecommand \@@startlink[1]{}%
\providecommand \@@endlink[0]{}%
\providecommand \url  [0]{\begingroup\@sanitize@url \@url }%
\providecommand \@url [1]{\endgroup\@href {#1}{\urlprefix }}%
\providecommand \urlprefix  [0]{URL }%
\providecommand \Eprint [0]{\href }%
\providecommand \doibase [0]{http://dx.doi.org/}%
\providecommand \selectlanguage [0]{\@gobble}%
\providecommand \bibinfo  [0]{\@secondoftwo}%
\providecommand \bibfield  [0]{\@secondoftwo}%
\providecommand \translation [1]{[#1]}%
\providecommand \BibitemOpen [0]{}%
\providecommand \bibitemStop [0]{}%
\providecommand \bibitemNoStop [0]{.\EOS\space}%
\providecommand \EOS [0]{\spacefactor3000\relax}%
\providecommand \BibitemShut  [1]{\csname bibitem#1\endcsname}%
\let\auto@bib@innerbib\@empty
\bibitem [{\citenamefont {Uchida}\ \emph {et~al.}(2008)\citenamefont {Uchida},
  \citenamefont {Takahashi}, \citenamefont {Harii}, \citenamefont {Ieda},
  \citenamefont {Koshibae}, \citenamefont {Ando}, \citenamefont {Maekawa},\
  and\ \citenamefont {Saitoh}}]{uchida2008}%
  \BibitemOpen
  \bibfield  {author} {\bibinfo {author} {\bibfnamefont {K.}~\bibnamefont
  {Uchida}}, \bibinfo {author} {\bibfnamefont {S.}~\bibnamefont {Takahashi}},
  \bibinfo {author} {\bibfnamefont {K.}~\bibnamefont {Harii}}, \bibinfo
  {author} {\bibfnamefont {J.}~\bibnamefont {Ieda}}, \bibinfo {author}
  {\bibfnamefont {W.}~\bibnamefont {Koshibae}}, \bibinfo {author}
  {\bibfnamefont {K.}~\bibnamefont {Ando}}, \bibinfo {author} {\bibfnamefont
  {S.}~\bibnamefont {Maekawa}}, \ and\ \bibinfo {author} {\bibfnamefont
  {E.}~\bibnamefont {Saitoh}},\ }\bibfield  {title} {\enquote {\bibinfo {title}
  {Observation of the spin seebeck effect},}\ }\href {\doibase
  doi:10.1038/nature07321} {\bibfield  {journal} {\bibinfo  {journal} {Nature}\
  }\textbf {\bibinfo {volume} {455}},\ \bibinfo {pages} {778 EP --} (\bibinfo
  {year} {2008})}\BibitemShut {NoStop}%
\bibitem [{\citenamefont {Uchida}\ \emph
  {et~al.}(2010{\natexlab{a}})\citenamefont {Uchida}, \citenamefont {Adachi},
  \citenamefont {Ota}, \citenamefont {Nakayama}, \citenamefont {Maekawa},\ and\
  \citenamefont {Saitoh}}]{uchida2010b}%
  \BibitemOpen
  \bibfield  {author} {\bibinfo {author} {\bibfnamefont {Ken-ichi}\
  \bibnamefont {Uchida}}, \bibinfo {author} {\bibfnamefont {Hiroto}\
  \bibnamefont {Adachi}}, \bibinfo {author} {\bibfnamefont {Takeru}\
  \bibnamefont {Ota}}, \bibinfo {author} {\bibfnamefont {Hiroyasu}\
  \bibnamefont {Nakayama}}, \bibinfo {author} {\bibfnamefont {Sadamichi}\
  \bibnamefont {Maekawa}}, \ and\ \bibinfo {author} {\bibfnamefont {Eiji}\
  \bibnamefont {Saitoh}},\ }\bibfield  {title} {\enquote {\bibinfo {title}
  {Observation of longitudinal spin-seebeck effect in magnetic insulators},}\
  }\href {\doibase 10.1063/1.3507386} {\bibfield  {journal} {\bibinfo
  {journal} {Applied Physics Letters}\ }\textbf {\bibinfo {volume} {97}},\
  \bibinfo {pages} {172505} (\bibinfo {year} {2010}{\natexlab{a}})}\BibitemShut {NoStop}%
\bibitem [{\citenamefont {Uchida}\ \emph
  {et~al.}(2010{\natexlab{b}})\citenamefont {Uchida}, \citenamefont {Xiao},
  \citenamefont {Adachi}, \citenamefont {Ohe}, \citenamefont {Takahashi},
  \citenamefont {Ieda}, \citenamefont {Ota}, \citenamefont {Kajiwara},
  \citenamefont {Umezawa}, \citenamefont {Kawai}, \citenamefont {Bauer},
  \citenamefont {Maekawa},\ and\ \citenamefont {Saitoh}}]{uchida2010}%
  \BibitemOpen
  \bibfield  {author} {\bibinfo {author} {\bibfnamefont {K.}~\bibnamefont
  {Uchida}}, \bibinfo {author} {\bibfnamefont {J.}~\bibnamefont {Xiao}},
  \bibinfo {author} {\bibfnamefont {H.}~\bibnamefont {Adachi}}, \bibinfo
  {author} {\bibfnamefont {J.}~\bibnamefont {Ohe}}, \bibinfo {author}
  {\bibfnamefont {S.}~\bibnamefont {Takahashi}}, \bibinfo {author}
  {\bibfnamefont {J.}~\bibnamefont {Ieda}}, \bibinfo {author} {\bibfnamefont
  {T.}~\bibnamefont {Ota}}, \bibinfo {author} {\bibfnamefont {Y.}~\bibnamefont
  {Kajiwara}}, \bibinfo {author} {\bibfnamefont {H.}~\bibnamefont {Umezawa}},
  \bibinfo {author} {\bibfnamefont {H.}~\bibnamefont {Kawai}}, \bibinfo
  {author} {\bibfnamefont {G.~E.~W.}\ \bibnamefont {Bauer}}, \bibinfo {author}
  {\bibfnamefont {S.}~\bibnamefont {Maekawa}}, \ and\ \bibinfo {author}
  {\bibfnamefont {E.}~\bibnamefont {Saitoh}},\ }\bibfield  {title} {\enquote
  {\bibinfo {title} {Spin seebeck insulator},}\ }\href {\doibase
  doi:10.1038/nmat2856} {\bibfield  {journal} {\bibinfo  {journal} {Nature
  Materials}\ }\textbf {\bibinfo {volume} {9}},\ \bibinfo {pages} {894 EP --}
  (\bibinfo {year} {2010}{\natexlab{b}})}\BibitemShut {NoStop}%
\bibitem [{\citenamefont {S.~Maekawa}\ and\ \citenamefont
  {Kimura}(2012)}]{maekawa2011}%
  \BibitemOpen
  \bibinfo {editor} {\bibfnamefont {E.~Saitoh}\ \bibnamefont {S.~Maekawa},
  \bibfnamefont {S.~O.~Valenzuela}}\ and\ \bibinfo {editor} {\bibfnamefont
  {T.}~\bibnamefont {Kimura}},\ eds.,\ \href@noop {} {\emph {\bibinfo {title}
  {Spin Current}}}\ (\bibinfo  {publisher} {Oxford University Press},\ \bibinfo
  {year} {2012})\BibitemShut {NoStop}%
\bibitem [{\citenamefont {Bauer}\ \emph {et~al.}(2012)\citenamefont {Bauer},
  \citenamefont {Saitoh},\ and\ \citenamefont {van Wees}}]{bauer2012}%
  \BibitemOpen
  \bibfield  {author} {\bibinfo {author} {\bibfnamefont {Gerrit E.~W.}\
  \bibnamefont {Bauer}}, \bibinfo {author} {\bibfnamefont {Eiji}\ \bibnamefont
  {Saitoh}}, \ and\ \bibinfo {author} {\bibfnamefont {Bart~J.}\ \bibnamefont
  {van Wees}},\ }\bibfield  {title} {\enquote {\bibinfo {title} {Spin
  caloritronics},}\ }\href {\doibase doi:10.1038/nmat3301} {\bibfield
  {journal} {\bibinfo  {journal} {Nature Materials}\ }\textbf {\bibinfo
  {volume} {11}},\ \bibinfo {pages} {391 EP --} (\bibinfo {year}
  {2012})}\BibitemShut {NoStop}%
\bibitem [{\citenamefont {Ohnuma}\ \emph {et~al.}(2013)\citenamefont {Ohnuma},
  \citenamefont {Adachi}, \citenamefont {Saitoh},\ and\ \citenamefont
  {Maekawa}}]{ohnuma2013}%
  \BibitemOpen
  \bibfield  {author} {\bibinfo {author} {\bibfnamefont {Yuichi}\ \bibnamefont
  {Ohnuma}}, \bibinfo {author} {\bibfnamefont {Hiroto}\ \bibnamefont {Adachi}},
  \bibinfo {author} {\bibfnamefont {Eiji}\ \bibnamefont {Saitoh}}, \ and\
  \bibinfo {author} {\bibfnamefont {Sadamichi}\ \bibnamefont {Maekawa}},\
  }\bibfield  {title} {\enquote {\bibinfo {title} {Spin seebeck effect in
  antiferromagnets and compensated ferrimagnets},}\ }\href {\doibase
  10.1103/PhysRevB.87.014423} {\bibfield  {journal} {\bibinfo  {journal} {Phys.
  Rev. B}\ }\textbf {\bibinfo {volume} {87}},\ \bibinfo {pages} {014423}
  (\bibinfo {year} {2013})}\BibitemShut {NoStop}%
\bibitem [{\citenamefont {Kikkawa}\ \emph {et~al.}(2013)\citenamefont
  {Kikkawa}, \citenamefont {Uchida}, \citenamefont {Shiomi}, \citenamefont
  {Qiu}, \citenamefont {Hou}, \citenamefont {Tian}, \citenamefont {Nakayama},
  \citenamefont {Jin},\ and\ \citenamefont {Saitoh}}]{kikkawa2013}%
  \BibitemOpen
  \bibfield  {author} {\bibinfo {author} {\bibfnamefont {T.}~\bibnamefont
  {Kikkawa}}, \bibinfo {author} {\bibfnamefont {K.}~\bibnamefont {Uchida}},
  \bibinfo {author} {\bibfnamefont {Y.}~\bibnamefont {Shiomi}}, \bibinfo
  {author} {\bibfnamefont {Z.}~\bibnamefont {Qiu}}, \bibinfo {author}
  {\bibfnamefont {D.}~\bibnamefont {Hou}}, \bibinfo {author} {\bibfnamefont
  {D.}~\bibnamefont {Tian}}, \bibinfo {author} {\bibfnamefont {H.}~\bibnamefont
  {Nakayama}}, \bibinfo {author} {\bibfnamefont {X.-F.}\ \bibnamefont {Jin}}, \
  and\ \bibinfo {author} {\bibfnamefont {E.}~\bibnamefont {Saitoh}},\
  }\bibfield  {title} {\enquote {\bibinfo {title} {Longitudinal spin seebeck
  effect free from the proximity nernst effect},}\ }\href {\doibase
  10.1103/PhysRevLett.110.067207} {\bibfield  {journal} {\bibinfo  {journal}
  {Phys. Rev. Lett.}\ }\textbf {\bibinfo {volume} {110}},\ \bibinfo {pages}
  {067207} (\bibinfo {year} {2013})}\BibitemShut {NoStop}%
\bibitem [{\citenamefont {Chumak}\ \emph {et~al.}(2015)\citenamefont {Chumak},
  \citenamefont {Vasyuchka}, \citenamefont {Serga},\ and\ \citenamefont
  {Hillebrands}}]{chumak2015}%
  \BibitemOpen
  \bibfield  {author} {\bibinfo {author} {\bibfnamefont {A.~V.}\ \bibnamefont
  {Chumak}}, \bibinfo {author} {\bibfnamefont {V.~I.}\ \bibnamefont
  {Vasyuchka}}, \bibinfo {author} {\bibfnamefont {A.~A.}\ \bibnamefont
  {Serga}}, \ and\ \bibinfo {author} {\bibfnamefont {B.}~\bibnamefont
  {Hillebrands}},\ }\bibfield  {title} {\enquote {\bibinfo {title} {Magnon
  spintronics},}\ }\href {\doibase doi:10.1038/nphys3347} {\bibfield  {journal}
  {\bibinfo  {journal} {Nat Phys}\ }\textbf {\bibinfo {volume} {11}},\ \bibinfo
  {pages} {453--461} (\bibinfo {year} {2015})}\BibitemShut {NoStop}%
\bibitem [{\citenamefont {Cornelissen}\ \emph {et~al.}(2015)\citenamefont
  {Cornelissen}, \citenamefont {Liu}, \citenamefont {Duine}, \citenamefont
  {Youssef},\ and\ \citenamefont {van Wees}}]{cornelissen2015}%
  \BibitemOpen
  \bibfield  {author} {\bibinfo {author} {\bibfnamefont {L.~J.}\ \bibnamefont
  {Cornelissen}}, \bibinfo {author} {\bibfnamefont {J.}~\bibnamefont {Liu}},
  \bibinfo {author} {\bibfnamefont {R.~A.}\ \bibnamefont {Duine}}, \bibinfo
  {author} {\bibfnamefont {J.~Ben}\ \bibnamefont {Youssef}}, \ and\ \bibinfo
  {author} {\bibfnamefont {B.~J.}\ \bibnamefont {van Wees}},\ }\bibfield
  {title} {\enquote {\bibinfo {title} {Long-distance transport of magnon spin
  information in a magnetic insulator at room temperature},}\ }\href {\doibase
  10.1038/nphys3465} {\bibfield  {journal} {\bibinfo  {journal} {Nat Phys}\
  }\textbf {\bibinfo {volume} {11}},\ \bibinfo {pages} {1022--1026} (\bibinfo
  {year} {2015})}\BibitemShut {NoStop}%
\bibitem [{\citenamefont {Chen}\ \emph {et~al.}(2013)\citenamefont {Chen},
  \citenamefont {Sun}, \citenamefont {Wang},\ and\ \citenamefont
  {Xie}}]{chen2013}%
  \BibitemOpen
  \bibfield  {author} {\bibinfo {author} {\bibfnamefont {Chui-Zhen}\
  \bibnamefont {Chen}}, \bibinfo {author} {\bibfnamefont {Qing-feng}\
  \bibnamefont {Sun}}, \bibinfo {author} {\bibfnamefont {Fa}~\bibnamefont
  {Wang}}, \ and\ \bibinfo {author} {\bibfnamefont {X.~C.}\ \bibnamefont
  {Xie}},\ }\bibfield  {title} {\enquote {\bibinfo {title} {Detection of
  spinons via spin transport},}\ }\href {\doibase 10.1103/PhysRevB.88.041405}
  {\bibfield  {journal} {\bibinfo  {journal} {Phys. Rev. B}\ }\textbf {\bibinfo
  {volume} {88}},\ \bibinfo {pages} {041405} (\bibinfo {year}
  {2013})}\BibitemShut {NoStop}%
\bibitem [{\citenamefont {Katsura}\ \emph {et~al.}(2010)\citenamefont
  {Katsura}, \citenamefont {Nagaosa},\ and\ \citenamefont
  {Lee}}]{kastsura2010}%
  \BibitemOpen
  \bibfield  {author} {\bibinfo {author} {\bibfnamefont {Hosho}\ \bibnamefont
  {Katsura}}, \bibinfo {author} {\bibfnamefont {Naoto}\ \bibnamefont
  {Nagaosa}}, \ and\ \bibinfo {author} {\bibfnamefont {Patrick~A.}\
  \bibnamefont {Lee}},\ }\bibfield  {title} {\enquote {\bibinfo {title} {Theory
  of the thermal hall effect in quantum magnets},}\ }\href {\doibase
  10.1103/PhysRevLett.104.066403} {\bibfield  {journal} {\bibinfo  {journal}
  {Phys. Rev. Lett.}\ }\textbf {\bibinfo {volume} {104}},\ \bibinfo {pages}
  {066403} (\bibinfo {year} {2010})}\BibitemShut {NoStop}%
\bibitem [{\citenamefont {Onose}\ \emph {et~al.}(2010)\citenamefont {Onose},
  \citenamefont {Ideue}, \citenamefont {Katsura}, \citenamefont {Shiomi},
  \citenamefont {Nagaosa},\ and\ \citenamefont {Tokura}}]{onose2010}%
  \BibitemOpen
  \bibfield  {author} {\bibinfo {author} {\bibfnamefont {Y.}~\bibnamefont
  {Onose}}, \bibinfo {author} {\bibfnamefont {T.}~\bibnamefont {Ideue}},
  \bibinfo {author} {\bibfnamefont {H.}~\bibnamefont {Katsura}}, \bibinfo
  {author} {\bibfnamefont {Y.}~\bibnamefont {Shiomi}}, \bibinfo {author}
  {\bibfnamefont {N.}~\bibnamefont {Nagaosa}}, \ and\ \bibinfo {author}
  {\bibfnamefont {Y.}~\bibnamefont {Tokura}},\ }\bibfield  {title} {\enquote
  {\bibinfo {title} {Observation of the magnon hall effect},}\ }\href {\doibase
  10.1126/science.1188260} {\bibfield  {journal} {\bibinfo  {journal}
  {Science}\ }\textbf {\bibinfo {volume} {329}},\ \bibinfo {pages} {297--299}
  (\bibinfo {year} {2010})} \BibitemShut
  {NoStop}%
\bibitem [{\citenamefont {Hirschberger}\ \emph
  {et~al.}(2015{\natexlab{a}})\citenamefont {Hirschberger}, \citenamefont
  {Chisnell}, \citenamefont {Lee},\ and\ \citenamefont
  {Ong}}]{hirschberger2015a}%
  \BibitemOpen
  \bibfield  {author} {\bibinfo {author} {\bibfnamefont {Max}\ \bibnamefont
  {Hirschberger}}, \bibinfo {author} {\bibfnamefont {Robin}\ \bibnamefont
  {Chisnell}}, \bibinfo {author} {\bibfnamefont {Young~S.}\ \bibnamefont
  {Lee}}, \ and\ \bibinfo {author} {\bibfnamefont {N.~P.}\ \bibnamefont
  {Ong}},\ }\bibfield  {title} {\enquote {\bibinfo {title} {Thermal hall effect
  of spin excitations in a kagome magnet},}\ }\href {\doibase
  10.1103/PhysRevLett.115.106603} {\bibfield  {journal} {\bibinfo  {journal}
  {Phys. Rev. Lett.}\ }\textbf {\bibinfo {volume} {115}},\ \bibinfo {pages}
  {106603} (\bibinfo {year} {2015}{\natexlab{a}})}\BibitemShut {NoStop}%
\bibitem [{\citenamefont {Hirschberger}\ \emph
  {et~al.}(2015{\natexlab{b}})\citenamefont {Hirschberger}, \citenamefont
  {Krizan}, \citenamefont {Cava},\ and\ \citenamefont
  {Ong}}]{hirschberger2015b}%
  \BibitemOpen
  \bibfield  {author} {\bibinfo {author} {\bibfnamefont {Max}\ \bibnamefont
  {Hirschberger}}, \bibinfo {author} {\bibfnamefont {Jason~W.}\ \bibnamefont
  {Krizan}}, \bibinfo {author} {\bibfnamefont {R.~J.}\ \bibnamefont {Cava}}, \
  and\ \bibinfo {author} {\bibfnamefont {N.~P.}\ \bibnamefont {Ong}},\
  }\bibfield  {title} {\enquote {\bibinfo {title} {Large thermal hall
  conductivity of neutral spin excitations in a frustrated quantum magnet},}\
  }\href {\doibase 10.1126/science.1257340} {\bibfield  {journal} {\bibinfo
  {journal} {Science}\ }\textbf {\bibinfo {volume} {348}},\ \bibinfo {pages}
  {106--109} (\bibinfo {year} {2015}{\natexlab{b}})} \BibitemShut
  {NoStop}%
\bibitem [{\citenamefont {Ideue}\ \emph {et~al.}(2017)\citenamefont {Ideue},
  \citenamefont {Kurumaji}, \citenamefont {Ishiwata},\ and\ \citenamefont
  {Tokura}}]{ideue2017}%
  \BibitemOpen
  \bibfield  {author} {\bibinfo {author} {\bibfnamefont {T.}~\bibnamefont
  {Ideue}}, \bibinfo {author} {\bibfnamefont {T.}~\bibnamefont {Kurumaji}},
  \bibinfo {author} {\bibfnamefont {S.}~\bibnamefont {Ishiwata}}, \ and\
  \bibinfo {author} {\bibfnamefont {Y.}~\bibnamefont {Tokura}},\ }\bibfield
  {title} {\enquote {\bibinfo {title} {Giant thermal hall effect in
  multiferroics},}\ }\href {http://dx.doi.org/10.1038/nmat4905} {\bibfield
  {journal} {\bibinfo  {journal} {Nature Materials}\ }\textbf {\bibinfo
  {volume} {16}},\ \bibinfo {pages} {797 EP --} (\bibinfo {year}
  {2017})}\BibitemShut {NoStop}%
\bibitem [{\citenamefont {Matsumoto}\ and\ \citenamefont
  {Murakami}(2011{\natexlab{a}})}]{matsumoto2011a}%
  \BibitemOpen
  \bibfield  {author} {\bibinfo {author} {\bibfnamefont {Ryo}\ \bibnamefont
  {Matsumoto}}\ and\ \bibinfo {author} {\bibfnamefont {Shuichi}\ \bibnamefont
  {Murakami}},\ }\bibfield  {title} {\enquote {\bibinfo {title} {Theoretical
  prediction of a rotating magnon wave packet in ferromagnets},}\ }\href
  {\doibase 10.1103/PhysRevLett.106.197202} {\bibfield  {journal} {\bibinfo
  {journal} {Phys Rev Lett}\ }\textbf {\bibinfo {volume} {106}},\ \bibinfo
  {pages} {197202} (\bibinfo {year} {2011}{\natexlab{a}})}\BibitemShut
  {NoStop}%
\bibitem [{\citenamefont {Matsumoto}\ and\ \citenamefont
  {Murakami}(2011{\natexlab{b}})}]{matsumoto2011b}%
  \BibitemOpen
  \bibfield  {author} {\bibinfo {author} {\bibfnamefont {Ryo}\ \bibnamefont
  {Matsumoto}}\ and\ \bibinfo {author} {\bibfnamefont {Shuichi}\ \bibnamefont
  {Murakami}},\ }\bibfield  {title} {\enquote {\bibinfo {title} {Rotational
  motion of magnons and the thermal hall effect},}\ }\href {\doibase
  10.1103/PhysRevB.84.184406} {\bibfield  {journal} {\bibinfo  {journal} {Phys.
  Rev. B}\ }\textbf {\bibinfo {volume} {84}},\ \bibinfo {pages} {184406}
  (\bibinfo {year} {2011}{\natexlab{b}})}\BibitemShut {NoStop}%
\bibitem [{\citenamefont {Zhang}\ \emph {et~al.}(2013)\citenamefont {Zhang},
  \citenamefont {Ren}, \citenamefont {Wang},\ and\ \citenamefont
  {Li}}]{zhang2013}%
  \BibitemOpen
  \bibfield  {author} {\bibinfo {author} {\bibfnamefont {Lifa}\ \bibnamefont
  {Zhang}}, \bibinfo {author} {\bibfnamefont {Jie}\ \bibnamefont {Ren}},
  \bibinfo {author} {\bibfnamefont {Jian-Sheng}\ \bibnamefont {Wang}}, \ and\
  \bibinfo {author} {\bibfnamefont {Baowen}\ \bibnamefont {Li}},\ }\bibfield
  {title} {\enquote {\bibinfo {title} {Topological magnon insulator in
  insulating ferromagnet},}\ }\href {\doibase 10.1103/PhysRevB.87.144101}
  {\bibfield  {journal} {\bibinfo  {journal} {Phys. Rev. B}\ }\textbf {\bibinfo
  {volume} {87}},\ \bibinfo {pages} {144101} (\bibinfo {year}
  {2013})}\BibitemShut {NoStop}%
\bibitem [{\citenamefont {Shindou}\ \emph {et~al.}(2013)\citenamefont
  {Shindou}, \citenamefont {Matsumoto}, \citenamefont {Murakami},\ and\
  \citenamefont {Ohe}}]{shindou2013}%
  \BibitemOpen
  \bibfield  {author} {\bibinfo {author} {\bibfnamefont {Ryuichi}\ \bibnamefont
  {Shindou}}, \bibinfo {author} {\bibfnamefont {Ryo}\ \bibnamefont
  {Matsumoto}}, \bibinfo {author} {\bibfnamefont {Shuichi}\ \bibnamefont
  {Murakami}}, \ and\ \bibinfo {author} {\bibfnamefont {Jun-ichiro}\
  \bibnamefont {Ohe}},\ }\bibfield  {title} {\enquote {\bibinfo {title}
  {Topological chiral magnonic edge mode in a magnonic crystal},}\ }\href
  {\doibase 10.1103/PhysRevB.87.174427} {\bibfield  {journal} {\bibinfo
  {journal} {Phys. Rev. B}\ }\textbf {\bibinfo {volume} {87}},\ \bibinfo
  {pages} {174427} (\bibinfo {year} {2013})}\BibitemShut {NoStop}%
\bibitem [{\citenamefont {Lee}\ \emph {et~al.}(2015)\citenamefont {Lee},
  \citenamefont {Han},\ and\ \citenamefont {Lee}}]{Lee2015}%
  \BibitemOpen
  \bibfield  {author} {\bibinfo {author} {\bibfnamefont {Hyunyong}\
  \bibnamefont {Lee}}, \bibinfo {author} {\bibfnamefont {Jung~Hoon}\
  \bibnamefont {Han}}, \ and\ \bibinfo {author} {\bibfnamefont {Patrick~A.}\
  \bibnamefont {Lee}},\ }\bibfield  {title} {\enquote {\bibinfo {title}
  {Thermal hall effect of spins in a paramagnet},}\ }\href {\doibase
  10.1103/PhysRevB.91.125413} {\bibfield  {journal} {\bibinfo  {journal} {Phys.
  Rev. B}\ }\textbf {\bibinfo {volume} {91}},\ \bibinfo {pages} {125413}
  (\bibinfo {year} {2015})}\BibitemShut {NoStop}%
\bibitem [{\citenamefont {Han}\ and\ \citenamefont {Lee}(2016)}]{han2016}%
  \BibitemOpen
  \bibfield  {author} {\bibinfo {author} {\bibfnamefont {Jung~Hoon}\
  \bibnamefont {Han}}\ and\ \bibinfo {author} {\bibfnamefont {Hyunyong}\
  \bibnamefont {Lee}},\ }\bibfield  {title} {\enquote {\bibinfo {title} {Spin
  chirality and hall-like transport phenomena of spin excitations},}\
  }\bibfield  {booktitle} {\emph {\bibinfo {booktitle} {Journal of the Physical
  Society of Japan}},\ }\href {\doibase 10.7566/JPSJ.86.011007} {\bibfield
  {journal} {\bibinfo  {journal} {Journal of the Physical Society of Japan}\
  }\textbf {\bibinfo {volume} {86}},\ \bibinfo {pages} {011007} (\bibinfo
  {year} {2016})}\BibitemShut {NoStop}%
\bibitem [{\citenamefont {Owerre}(2016{\natexlab{a}})}]{owerre2016}%
  \BibitemOpen
  \bibfield  {author} {\bibinfo {author} {\bibfnamefont {S.~A.}\ \bibnamefont
  {Owerre}},\ }\bibfield  {title} {\enquote {\bibinfo {title} {Topological
  honeycomb magnon hall effect: A calculation of thermal hall conductivity of
  magnetic spin excitations},}\ }\href {\doibase 10.1063/1.4959815} {\bibfield
  {journal} {\bibinfo  {journal} {Journal of Applied Physics}\ }\textbf
  {\bibinfo {volume} {120}},\ \bibinfo {pages} {043903} (\bibinfo {year}
  {2016}{\natexlab{a}})}\BibitemShut {NoStop}%
\bibitem [{\citenamefont {Owerre}(2016{\natexlab{b}})}]{owerre2016b}%
  \BibitemOpen
  \bibfield  {author} {\bibinfo {author} {\bibfnamefont {S.~A.}\ \bibnamefont
  {Owerre}},\ }\bibfield  {title} {\enquote {\bibinfo {title} {Magnon hall
  effect in ab-stacked bilayer honeycomb quantum magnets},}\ }\href {\doibase
  10.1103/PhysRevB.94.094405} {\bibfield  {journal} {\bibinfo  {journal} {Phys.
  Rev. B}\ }\textbf {\bibinfo {volume} {94}},\ \bibinfo {pages} {094405}
  (\bibinfo {year} {2016}{\natexlab{b}})}\BibitemShut {NoStop}%
\bibitem [{\citenamefont {Owerre}(2017)}]{owerre2017}%
  \BibitemOpen
  \bibfield  {author} {\bibinfo {author} {\bibfnamefont {S.~A.}\ \bibnamefont
  {Owerre}},\ }\bibfield  {title} {\enquote {\bibinfo {title} {Topological
  thermal hall effect in frustrated kagome antiferromagnets},}\ }\href
  {\doibase 10.1103/PhysRevB.95.014422} {\bibfield  {journal} {\bibinfo
  {journal} {Phys. Rev. B}\ }\textbf {\bibinfo {volume} {95}},\ \bibinfo
  {pages} {014422} (\bibinfo {year} {2017})}\BibitemShut {NoStop}%
\bibitem [{\citenamefont {Cheng}\ \emph {et~al.}(2016)\citenamefont {Cheng},
  \citenamefont {Okamoto},\ and\ \citenamefont {Xiao}}]{cheng2016}%
  \BibitemOpen
  \bibfield  {author} {\bibinfo {author} {\bibfnamefont {Ran}\ \bibnamefont
  {Cheng}}, \bibinfo {author} {\bibfnamefont {Satoshi}\ \bibnamefont
  {Okamoto}}, \ and\ \bibinfo {author} {\bibfnamefont {Di}~\bibnamefont
  {Xiao}},\ }\bibfield  {title} {\enquote {\bibinfo {title} {Spin nernst effect
  of magnons in collinear antiferromagnets},}\ }\href {\doibase
  10.1103/PhysRevLett.117.217202} {\bibfield  {journal} {\bibinfo  {journal}
  {Phys. Rev. Lett.}\ }\textbf {\bibinfo {volume} {117}},\ \bibinfo {pages}
  {217202} (\bibinfo {year} {2016})}\BibitemShut {NoStop}%
\bibitem [{\citenamefont {Zyuzin}\ and\ \citenamefont
  {Kovalev}(2016)}]{zyuzin2016}%
  \BibitemOpen
  \bibfield  {author} {\bibinfo {author} {\bibfnamefont {Vladimir~A.}\
  \bibnamefont {Zyuzin}}\ and\ \bibinfo {author} {\bibfnamefont {Alexey~A.}\
  \bibnamefont {Kovalev}},\ }\bibfield  {title} {\enquote {\bibinfo {title}
  {Magnon spin nernst effect in antiferromagnets},}\ }\href {\doibase
  10.1103/PhysRevLett.117.217203} {\bibfield  {journal} {\bibinfo  {journal}
  {Phys. Rev. Lett.}\ }\textbf {\bibinfo {volume} {117}},\ \bibinfo {pages}
  {217203} (\bibinfo {year} {2016})}\BibitemShut {NoStop}%
\bibitem [{\citenamefont {Lee}\ \emph {et~al.}(2018)\citenamefont {Lee},
  \citenamefont {Chung}, \citenamefont {Park},\ and\ \citenamefont
  {Park}}]{lee2018}%
  \BibitemOpen
  \bibfield  {author} {\bibinfo {author} {\bibfnamefont {Ki~Hoon}\ \bibnamefont
  {Lee}}, \bibinfo {author} {\bibfnamefont {Suk~Bum}\ \bibnamefont {Chung}},
  \bibinfo {author} {\bibfnamefont {Kisoo}\ \bibnamefont {Park}}, \ and\
  \bibinfo {author} {\bibfnamefont {Je-Geun}\ \bibnamefont {Park}},\ }\bibfield
   {title} {\enquote {\bibinfo {title} {Magnonic quantum spin hall state in the
  zigzag and stripe phases of the antiferromagnetic honeycomb lattice},}\
  }\href {\doibase 10.1103/PhysRevB.97.180401} {\bibfield  {journal} {\bibinfo
  {journal} {Phys. Rev. B}\ }\textbf {\bibinfo {volume} {97}},\ \bibinfo
  {pages} {180401} (\bibinfo {year} {2018})}\BibitemShut {NoStop}%
\bibitem [{\citenamefont {Kim}\ \emph {et~al.}(2016)\citenamefont {Kim},
  \citenamefont {Ochoa}, \citenamefont {Zarzuela},\ and\ \citenamefont
  {Tserkovnyak}}]{kim2016}%
  \BibitemOpen
  \bibfield  {author} {\bibinfo {author} {\bibfnamefont {Se~Kwon}\ \bibnamefont
  {Kim}}, \bibinfo {author} {\bibfnamefont {H\'ector}\ \bibnamefont {Ochoa}},
  \bibinfo {author} {\bibfnamefont {Ricardo}\ \bibnamefont {Zarzuela}}, \ and\
  \bibinfo {author} {\bibfnamefont {Yaroslav}\ \bibnamefont {Tserkovnyak}},\
  }\bibfield  {title} {\enquote {\bibinfo {title} {Realization of the
  haldane-kane-mele model in a system of localized spins},}\ }\href {\doibase
  10.1103/PhysRevLett.117.227201} {\bibfield  {journal} {\bibinfo  {journal}
  {Phys. Rev. Lett.}\ }\textbf {\bibinfo {volume} {117}},\ \bibinfo {pages}
  {227201} (\bibinfo {year} {2016})}\BibitemShut {NoStop}%
\bibitem [{\citenamefont {Shiomi}\ \emph {et~al.}(2017)\citenamefont {Shiomi},
  \citenamefont {Takashima},\ and\ \citenamefont {Saitoh}}]{saitoh2017b}%
  \BibitemOpen
  \bibfield  {author} {\bibinfo {author} {\bibfnamefont {Y.}~\bibnamefont
  {Shiomi}}, \bibinfo {author} {\bibfnamefont {R.}~\bibnamefont {Takashima}}, \
  and\ \bibinfo {author} {\bibfnamefont {E.}~\bibnamefont {Saitoh}},\
  }\bibfield  {title} {\enquote {\bibinfo {title} {Experimental evidence
  consistent with a magnon nernst effect in the antiferromagnetic insulator
  ${\mathrm{mnps}}_{3}$},}\ }\href {\doibase 10.1103/PhysRevB.96.134425}
  {\bibfield  {journal} {\bibinfo  {journal} {Phys. Rev. B}\ }\textbf {\bibinfo
  {volume} {96}},\ \bibinfo {pages} {134425} (\bibinfo {year}
  {2017})}\BibitemShut {NoStop}%
\bibitem [{\citenamefont {Dzyaloshinsky}(1958)}]{Dzyaloshinsky1958}%
  \BibitemOpen
  \bibfield  {author} {\bibinfo {author} {\bibfnamefont {I.}~\bibnamefont
  {Dzyaloshinsky}},\ }\bibfield  {title} {\enquote {\bibinfo {title} {A
  thermodynamic theory of ``weak'' ferromagnetism of antiferromagnetics},}\
  }\href {\doibase https://doi.org/10.1016/0022-3697(58)90076-3} {\bibfield
  {journal} {\bibinfo  {journal} {Journal of Physics and Chemistry of Solids}\
  }\textbf {\bibinfo {volume} {4}},\ \bibinfo {pages} {241 -- 255} (\bibinfo
  {year} {1958})}\BibitemShut {NoStop}%
\bibitem [{\citenamefont {Moriya}(1960)}]{moriya1960}%
  \BibitemOpen
  \bibfield  {author} {\bibinfo {author} {\bibfnamefont {T\^oru}\ \bibnamefont
  {Moriya}},\ }\bibfield  {title} {\enquote {\bibinfo {title} {Anisotropic
  superexchange interaction and weak ferromagnetism},}\ }\href {\doibase
  10.1103/PhysRev.120.91} {\bibfield  {journal} {\bibinfo  {journal} {Phys.
  Rev.}\ }\textbf {\bibinfo {volume} {120}},\ \bibinfo {pages} {91--98}
  (\bibinfo {year} {1960})}\BibitemShut {NoStop}%
\bibitem [{\citenamefont {Timm}\ and\ \citenamefont {Jensen}(2000)}]{timm2000}%
  \BibitemOpen
  \bibfield  {author} {\bibinfo {author} {\bibfnamefont {Carsten}\ \bibnamefont
  {Timm}}\ and\ \bibinfo {author} {\bibfnamefont {P.~J.}\ \bibnamefont
  {Jensen}},\ }\bibfield  {title} {\enquote {\bibinfo {title} {Schwinger boson
  theory of anisotropic ferromagnetic ultrathin films},}\ }\href {\doibase
  10.1103/PhysRevB.62.5634} {\bibfield  {journal} {\bibinfo  {journal} {Phys.
  Rev. B}\ }\textbf {\bibinfo {volume} {62}},\ \bibinfo {pages} {5634--5646}
  (\bibinfo {year} {2000})}\BibitemShut {NoStop}%
\bibitem [{\citenamefont {Auerbach}(1994)}]{auerbach209}%
  \BibitemOpen
  \bibfield  {author} {\bibinfo {author} {\bibfnamefont {Assa}\ \bibnamefont
  {Auerbach}},\ }\href {\doibase 10.1007/978-1-4612-0869-3} {\emph {\bibinfo
  {title} {Interacting Electrons and Quantum Magnetism}}}\ (\bibinfo
  {publisher} {Springer-Verlag New York},\ \bibinfo {year} {1994})\BibitemShut
  {NoStop}%
\bibitem [{\citenamefont {Sarker}\ \emph {et~al.}(1989)\citenamefont {Sarker},
  \citenamefont {Jayaprakash}, \citenamefont {Krishnamurthy},\ and\
  \citenamefont {Ma}}]{sarker1989}%
  \BibitemOpen
  \bibfield  {author} {\bibinfo {author} {\bibfnamefont {Sanjoy}\ \bibnamefont
  {Sarker}}, \bibinfo {author} {\bibfnamefont {C.}~\bibnamefont {Jayaprakash}},
  \bibinfo {author} {\bibfnamefont {H.~R.}\ \bibnamefont {Krishnamurthy}}, \
  and\ \bibinfo {author} {\bibfnamefont {Michael}\ \bibnamefont {Ma}},\
  }\bibfield  {title} {\enquote {\bibinfo {title} {Bosonic mean-field theory of
  quantum heisenberg spin systems: Bose condensation and magnetic order},}\
  }\href {\doibase 10.1103/PhysRevB.40.5028} {\bibfield  {journal} {\bibinfo
  {journal} {Phys. Rev. B}\ }\textbf {\bibinfo {volume} {40}},\ \bibinfo
  {pages} {5028--5035} (\bibinfo {year} {1989})}\BibitemShut {NoStop}%
\bibitem [{\citenamefont {Read}\ and\ \citenamefont
  {Sachdev}(1991)}]{read1991}%
  \BibitemOpen
  \bibfield  {author} {\bibinfo {author} {\bibfnamefont {N.}~\bibnamefont
  {Read}}\ and\ \bibinfo {author} {\bibfnamefont {Subir}\ \bibnamefont
  {Sachdev}},\ }\bibfield  {title} {\enquote {\bibinfo {title} {Large-n
  expansion for frustrated quantum antiferromagnets},}\ }\href {\doibase
  10.1103/PhysRevLett.66.1773} {\bibfield  {journal} {\bibinfo  {journal}
  {Phys. Rev. Lett.}\ }\textbf {\bibinfo {volume} {66}},\ \bibinfo {pages}
  {1773--1776} (\bibinfo {year} {1991})}\BibitemShut {NoStop}%
\bibitem [{\citenamefont {Zhou}\ \emph {et~al.}(2017)\citenamefont {Zhou},
  \citenamefont {Kanoda},\ and\ \citenamefont {Ng}}]{zhou2017}%
  \BibitemOpen
  \bibfield  {author} {\bibinfo {author} {\bibfnamefont {Yi}~\bibnamefont
  {Zhou}}, \bibinfo {author} {\bibfnamefont {Kazushi}\ \bibnamefont {Kanoda}},
  \ and\ \bibinfo {author} {\bibfnamefont {Tai-Kai}\ \bibnamefont {Ng}},\
  }\bibfield  {title} {\enquote {\bibinfo {title} {Quantum spin liquid
  states},}\ }\href {\doibase 10.1103/RevModPhys.89.025003} {\bibfield
  {journal} {\bibinfo  {journal} {Rev. Mod. Phys.}\ }\textbf {\bibinfo {volume}
  {89}},\ \bibinfo {pages} {025003} (\bibinfo {year} {2017})}\BibitemShut
  {NoStop}%
\bibitem [{\citenamefont {Nagaosa}\ \emph {et~al.}(2010)\citenamefont
  {Nagaosa}, \citenamefont {Sinova}, \citenamefont {Onoda}, \citenamefont
  {MacDonald},\ and\ \citenamefont {Ong}}]{nagaosa2010}%
  \BibitemOpen
  \bibfield  {author} {\bibinfo {author} {\bibfnamefont {Naoto}\ \bibnamefont
  {Nagaosa}}, \bibinfo {author} {\bibfnamefont {Jairo}\ \bibnamefont {Sinova}},
  \bibinfo {author} {\bibfnamefont {Shigeki}\ \bibnamefont {Onoda}}, \bibinfo
  {author} {\bibfnamefont {A.~H.}\ \bibnamefont {MacDonald}}, \ and\ \bibinfo
  {author} {\bibfnamefont {N.~P.}\ \bibnamefont {Ong}},\ }\bibfield  {title}
  {\enquote {\bibinfo {title} {Anomalous hall effect},}\ }\href {\doibase
  10.1103/RevModPhys.82.1539} {\bibfield  {journal} {\bibinfo  {journal} {Rev.
  Mod. Phys.}\ }\textbf {\bibinfo {volume} {82}},\ \bibinfo {pages}
  {1539--1592} (\bibinfo {year} {2010})}\BibitemShut {NoStop}%
\bibitem [{\citenamefont {Gusynin}\ \emph {et~al.}(2014)\citenamefont
  {Gusynin}, \citenamefont {Sharapov},\ and\ \citenamefont
  {Varlamov}}]{gusynin2014}%
  \BibitemOpen
  \bibfield  {author} {\bibinfo {author} {\bibfnamefont {V.~P.}\ \bibnamefont
  {Gusynin}}, \bibinfo {author} {\bibfnamefont {S.~G.}\ \bibnamefont
  {Sharapov}}, \ and\ \bibinfo {author} {\bibfnamefont {A.~A.}\ \bibnamefont
  {Varlamov}},\ }\bibfield  {title} {\enquote {\bibinfo {title} {Anomalous
  thermospin effect in the low-buckled dirac materials},}\ }\href {\doibase
  10.1103/PhysRevB.90.155107} {\bibfield  {journal} {\bibinfo  {journal} {Phys.
  Rev. B}\ }\textbf {\bibinfo {volume} {90}},\ \bibinfo {pages} {155107}
  (\bibinfo {year} {2014})}\BibitemShut {NoStop}%
\bibitem [{\citenamefont {Gusynin}\ \emph {et~al.}(2015)\citenamefont
  {Gusynin}, \citenamefont {Sharapov},\ and\ \citenamefont
  {Varlamov}}]{gusynin2015}%
  \BibitemOpen
  \bibfield  {author} {\bibinfo {author} {\bibfnamefont {V.~P.}\ \bibnamefont
  {Gusynin}}, \bibinfo {author} {\bibfnamefont {S.~G.}\ \bibnamefont
  {Sharapov}}, \ and\ \bibinfo {author} {\bibfnamefont {A.~A.}\ \bibnamefont
  {Varlamov}},\ }\bibfield  {title} {\enquote {\bibinfo {title} {Spin nernst
  effect and intrinsic magnetization in two-dimensional dirac materials},}\
  }\bibfield  {booktitle} {\emph {\bibinfo {booktitle} {Low Temperature
  Physics}},\ }\href {\doibase 10.1063/1.4919372} {\bibfield  {journal}
  {\bibinfo  {journal} {Low Temperature Physics}\ }\textbf {\bibinfo {volume}
  {41}},\ \bibinfo {pages} {342--352} (\bibinfo {year} {2015})}\BibitemShut
  {NoStop}%
\bibitem [{\citenamefont {Dyrda\l{}}\ \emph {et~al.}(2016)\citenamefont
  {Dyrda\l{}}, \citenamefont {Barna\ifmmode~\acute{s}\else \'{s}\fi{}},\ and\
  \citenamefont {Dugaev}}]{dyrda2016}%
  \BibitemOpen
  \bibfield  {author} {\bibinfo {author} {\bibfnamefont {A.}~\bibnamefont
  {Dyrda\l{}}}, \bibinfo {author} {\bibfnamefont {J.}~\bibnamefont
  {Barna\ifmmode~\acute{s}\else \'{s}\fi{}}}, \ and\ \bibinfo {author}
  {\bibfnamefont {V.~K.}\ \bibnamefont {Dugaev}},\ }\bibfield  {title}
  {\enquote {\bibinfo {title} {Spin hall and spin nernst effects in a
  two-dimensional electron gas with rashba spin-orbit interaction: Temperature
  dependence},}\ }\href {\doibase 10.1103/PhysRevB.94.035306} {\bibfield
  {journal} {\bibinfo  {journal} {Phys. Rev. B}\ }\textbf {\bibinfo {volume}
  {94}},\ \bibinfo {pages} {035306} (\bibinfo {year} {2016})}\BibitemShut
  {NoStop}%
\bibitem [{\citenamefont {Sivadas}\ \emph {et~al.}(2015)\citenamefont
  {Sivadas}, \citenamefont {Daniels}, \citenamefont {Swendsen}, \citenamefont
  {Okamoto},\ and\ \citenamefont {Xiao}}]{sivadas2015}%
  \BibitemOpen
  \bibfield  {author} {\bibinfo {author} {\bibfnamefont {Nikhil}\ \bibnamefont
  {Sivadas}}, \bibinfo {author} {\bibfnamefont {Matthew~W.}\ \bibnamefont
  {Daniels}}, \bibinfo {author} {\bibfnamefont {Robert~H.}\ \bibnamefont
  {Swendsen}}, \bibinfo {author} {\bibfnamefont {Satoshi}\ \bibnamefont
  {Okamoto}}, \ and\ \bibinfo {author} {\bibfnamefont {Di}~\bibnamefont
  {Xiao}},\ }\bibfield  {title} {\enquote {\bibinfo {title} {Magnetic ground
  state of semiconducting transition-metal trichalcogenide monolayers},}\
  }\href {\doibase 10.1103/PhysRevB.91.235425} {\bibfield  {journal} {\bibinfo
  {journal} {Phys. Rev. B}\ }\textbf {\bibinfo {volume} {91}},\ \bibinfo
  {pages} {235425} (\bibinfo {year} {2015})}\BibitemShut {NoStop}%
\bibitem [{\citenamefont {Huang}\ \emph {et~al.}(2017)\citenamefont {Huang},
  \citenamefont {Clark}, \citenamefont {Navarro-Moratalla}, \citenamefont
  {Klein}, \citenamefont {Cheng}, \citenamefont {Seyler}, \citenamefont
  {Zhong}, \citenamefont {Schmidgall}, \citenamefont {McGuire}, \citenamefont
  {Cobden}, \citenamefont {Yao}, \citenamefont {Xiao}, \citenamefont
  {Jarillo-Herrero},\ and\ \citenamefont {Xu}}]{huang2017}%
  \BibitemOpen
  \bibfield  {author} {\bibinfo {author} {\bibfnamefont {Bevin}\ \bibnamefont
  {Huang}}, \bibinfo {author} {\bibfnamefont {Genevieve}\ \bibnamefont
  {Clark}}, \bibinfo {author} {\bibfnamefont {Efr{\'e}n}\ \bibnamefont
  {Navarro-Moratalla}}, \bibinfo {author} {\bibfnamefont {Dahlia~R.}\
  \bibnamefont {Klein}}, \bibinfo {author} {\bibfnamefont {Ran}\ \bibnamefont
  {Cheng}}, \bibinfo {author} {\bibfnamefont {Kyle~L.}\ \bibnamefont {Seyler}},
  \bibinfo {author} {\bibfnamefont {Ding}\ \bibnamefont {Zhong}}, \bibinfo
  {author} {\bibfnamefont {Emma}\ \bibnamefont {Schmidgall}}, \bibinfo {author}
  {\bibfnamefont {Michael~A.}\ \bibnamefont {McGuire}}, \bibinfo {author}
  {\bibfnamefont {David~H.}\ \bibnamefont {Cobden}}, \bibinfo {author}
  {\bibfnamefont {Wang}\ \bibnamefont {Yao}}, \bibinfo {author} {\bibfnamefont
  {Di}~\bibnamefont {Xiao}}, \bibinfo {author} {\bibfnamefont {Pablo}\
  \bibnamefont {Jarillo-Herrero}}, \ and\ \bibinfo {author} {\bibfnamefont
  {Xiaodong}\ \bibnamefont {Xu}},\ }\bibfield  {title} {\enquote {\bibinfo
  {title} {Layer-dependent ferromagnetism in a van der waals crystal down to
  the monolayer limit},}\ }\href {http://dx.doi.org/10.1038/nature22391}
  {\bibfield  {journal} {\bibinfo  {journal} {Nature}\ }\textbf {\bibinfo
  {volume} {546}},\ \bibinfo {pages} {270 EP --} (\bibinfo {year}
  {2017})}\BibitemShut {NoStop}%
\end{thebibliography}

%

\end{document}